\documentclass[preprint,authoryear,12pt]{elsarticle}

\usepackage{amsfonts}
\usepackage{amssymb}
\usepackage{amsmath,epsfig}
\usepackage{fmtcount}
\usepackage{multirow}
\usepackage{array}
\usepackage{rotating}
\usepackage{fmtcount}
\usepackage{dsfont}
\usepackage{url}
\usepackage{threeparttable}
\usepackage{caption}
\usepackage{nomencl}
\usepackage{soul}
\usepackage{color}
\usepackage{algorithm}
\usepackage{algorithmic}
\usepackage{subfigure}
\usepackage{setspace}

\journal{Medical Image Analysis}

\begin{document}
\def\vector{\underline}
\def\matrix{}
\renewcommand{\vec}[1]{\mathbf{#1}}
\newcommand{\mat}[1]{\mathbf{#1}}
\begin{frontmatter}



\title{Compressed sensing for longitudinal MRI: An adaptive-weighted approach}

\author[technion]{L.~Weizman\corref{cor1}}
\ead{weizmanl@tx.technion.ac.il}
\author[technion]{Y.C.~Eldar}
\ead{yonina@ee.technion.ac.il}
\author[ichilov]{D.~Ben Bashat}
\ead{dafnab@tasmc.health.gov.il}
\cortext[cor1]{Corresponding author}
\address[technion]{Department of Electrical Engineering, Technion - Israel Institue of Technology, Israel}
\address[ichilov]{Tel Aviv Medical Center,  Tel Aviv University, Israel}

\begin{abstract}

{\bf Purpose:} Repeated brain MRI scans are  performed in many clinical scenarios, such as follow up of patients with tumors and therapy response assessment. In this paper, the authors show an approach to utilize former scans of the patient for the acceleration of repeated MRI scans.  

{\bf Methods:} The proposed approach utilizes the possible similarity of the repeated scans in longitudinal MRI studies. Since similarity is not guaranteed, sampling and reconstruction are adjusted during acquisition to match the actual similarity between the scans.
The baseline MR scan is utilized both in the sampling stage, via adaptive sampling, and in the reconstruction stage, with weighted reconstruction. In adaptive sampling, {\bf\emph k}-space sampling locations are optimized during acquisition. Weighted reconstruction uses the locations of the nonzero coefficients in the sparse domains as a prior in the recovery process. The approach was tested on 2D and 3D MRI scans of patients with brain tumors.

{\bf Results:} The longitudinal adaptive CS MRI (LACS-MRI) scheme provides reconstruction quality which outperforms other CS-based approaches for rapid MRI. Examples are shown on patients with brain tumors and demonstrate improved spatial resolution. Compared with data sampled at Nyquist rate, LACS-MRI exhibits Signal-to-Error Ratio (SER) of 24.8dB with undersampling factor of 16.6 in 3D MRI.

{\bf Conclusions:} The authors have presented a novel method for image reconstruction utilizing similarity of scans in longitudinal MRI studies, where possible. The proposed approach can play a major part and significantly reduce scanning time in many applications that consist of disease follow-up and monitoring of longitudinal changes in brain MRI.

\end{abstract}


\end{frontmatter}


\section{Introduction}
\label{Introduction}
Repeated brain MRI scans are  performed in many clinical scenarios, such as follow up of patients with tumors and therapy response assessment \citep{rees2009volumes,young2011longitudinal,weizman2012automatic,yip2014prior,weizman2014semiautomatic}. They constitute one of the most efficient tools to track pathology changes and to evaluate treatment efficacy in brain diseases. In many cases most of the imaging data of the repeated scan is already present in the former scan. In this paper we aim at exploiting this temporal similarity in MRI longitudinal studies for rapid MRI acquisition of the repeated scan.

The use of a reference image in medical image reconstruction is quite established and is popular in various imaging modalities. Examples include computed tomography \citep{chen2008prior,lauzier2012prior}, spectroscopic imaging \citep{hu1988slim}, imaging of contrast agent uptake \citep{van1993keyhole}, dynamic MRI \citep{jones1993k,liang1994efficient,korosec1996time,madore1999unaliasing,tsao2003k,mistretta2006highly}   and real time tracking of tumors \citep{yip2014prior},
 locally-focused MRI \citep{cao1995locally} and feature-recognizing MRI \citep{cao1993feature}. 
 
Since the introduction of Compressed Sensing (CS) \citep{candes2006compressive,donoho2006compressed,eldar2012compressed} to the field of MRI \citep{lustig2007sparse}, the use of a reference image has been exploited within CS, such as in rapid dynamic MRI, by exploiting temporal sparsity. Gamper et al.  perform randomly skipping phase-encoding lines in each dynamic frame to speed-up acquisition \citep{gamper2008compressed}. Liang et al. propose an iterative algorithm to detect the signal support in CS dynamic MRI \citep{liang2012k}. Zonooni and Kassim use the previous time-frame in dynamic MRI to weight the $\ell_1$ minimization in the CS reconstruction process \citep{zonoobi2013reconstruction}. Trzasko et al. exploit a pre-injection background image \citep{trzasko2011sparse} and Wu et al. utilize a constraining image to enhance  MR angiography (MRA), while incorporating CS and parallel imaging \citep{samsonov_ismrm_08_1}.

Samsonov et al.\citep{samsonov_ismrm} have suggested using a reference frame to speed-up MRI in longitudinal studies . Their approach, as well as most other approaches that exploit a reference image in MRI applications, rely on similarity between the reference scan and the current scan. The similarity assumption is indeed valid in cases where imaging data consists of many images acquired at a high frame rate. 

However, longitudinal MRI poses a different challenge due to the large time gaps between the scans. On the one hand, similarity across time points is not guaranteed, since in many cases we observe vast changes between scans due to pathology changes or surgical interventions. In addition, undersampling in the time domain is impractical, due to the demand for high quality reconstruction at each time point. Therefore, reconstruction errors in this single time-frame imaging modality cannot be compensated for by adjacent time-frames like in dynamic imaging. On the other hand, in cases where similarity between previous and current scans does exist, the high resolution former scan of the patient may constitute a very strong prior for the reconstruction of the repeated scan. Therefore, for the scenario of longitudinal MRI, we suggest exploiting the former scan in an adaptive manner. The similarity of the reference scan to the current scans is ``learned" during the acquisition process, leading to an iterative update of sampling and reconstruction accordingly. 

In the context of MRI, image reconstruction quality highly depends on the {\bf\emph k}-space sampling pattern \citep{tsai2000reduced,knoll2011adapted}. The concept of adaptive sampling (a.k.a ``adaptive sensing") suggests that samples are selected sequentially, where the choice of the next samples may depend on previously gathered information. This concept has been implemented previously, mainly for dynamic MRI \citep{panych1994dynamically,yoo1999real}, and was later extended and implemented in a CS framework \citep{seeger2010optimization,ravishankar2011adaptive,ravishankar2011mr}. In this work we utilize this concept and explore whether we can adaptively optimize the sampling pattern on-the-fly, based on partial reconstruction results from previously acquired samples in longitudinal MRI. 

A substantial body of mathematical theory has recently been published establishing the basic principles of adaptive sampling of sparse signals \citep{haupt2011distilled,wei2013,chen2014completing}. According to these mathematical results, reconstruction from samples selected sequentially, based on partial reconstruction results is significantly improved versus reconstruction from non-adaptive (deterministic or random) samples. 

Knowledge of the former scan can be advantageously used not only to design an adaptive sampling pattern, but also to improve image reconstruction from sampled data. This improvement can be obtained via the definition of regularization weights in the reconstruction optimization problem \citep{candes2008enhancing,haldar2008anatomically,vaswani2010modified}. We apply this approach, coined ``weighted reconstruction", for the scenario of longitudinal MRI. This weighting mechanism allows to relax or enforce the demand for sparsity according to the level of similarity between the current scan and the reference scan.  

In this paper we develop a framework for CS longitudinal MRI, by employing the two well established approaches above in parallel: adaptive sampling and weighted reconstruction. In this way, we exploit the prior scan of the patient to reconstruct the repeated scan from highly undersampled {\bf\emph k}-space data. To keep the discussion as simple as possible, we focus on Cartesian sampling for brain MRI. 

The novelty of this paper lies in the unique implementation of weighted reconstruction and adaptive sampling for the scenario of longitudinal studies, where the temporal similarity is not taken for granted. Unlike traditional CS MRI approaches that utilize prior constraints, in our approach the temporal similarity assumption is continuously examined, and the sampling and reconstruction algorithms are updated accordingly. Experimental results exhibit the superiority of the proposed method regardless of the validity of the temporal similarity assumption in the examined cases.  

This paper is organized as follows. Section \ref{methodology} presents the theory of weighted reconstruction and adaptive sampling and their implementation for longitudinal MRI. Section \ref{results} describes the
experimental results. Section \ref{discussion} discusses practical issues related to the implementation of the method in real time applications; Section \ref{conclusions} concludes by highlighting the key findings of the research.

\section{Method}
\label{methodology}
\subsection{Summary of Compressed Sensing MRI}
The application of CS for rapid MRI \citep{lustig2007sparse} exploits the fact that MRI scans are typically sparse in some transform domain, which is incoherent with the sampling domain. Nonlinear reconstruction is then used to enforce both sparsity of the image representation and consistency with the acquired data. A typical formulation of CS MRI recovery aims to solve the following constrained optimization problem:
\begin{equation}
\begin{aligned}
& \underset{\vec{x}}{\text{min}}
& & \|\mat{\Psi}\vec{x}\|_1 \ \ \text{s.t.}
&& \|\mat{F}_u\vec{x}-\vec{y}\|_2<\epsilon
\end{aligned}
\label{eq1}
\end{equation}
where $\vec{x}\in \mathbb{C}^{N}$ is the $N$-pixel complex image to be reconstructed, represented as a vector, $\vec{y}\in \mathbb{C}^{M}$ represents the {\bf\emph k}-space measurements, $\mat{F}_u$ is the undersampled Fourier transform operator, $\mat{\Psi}$ is a sparsifying transform 
operator and $\epsilon$ controls the fidelity of the reconstruction to the measured data. 

This fundamental CS MRI formulation is the basis for many MRI reconstruction applications, where the sparse transform domain varies depending on the particular setting. In MR angiography, where images are truly sparse, finite-differences is used as a sparsifying transform. In dynamic MRI, the difference between adjacent time frames is sparse \citep{lang2008accelerating,jung2009k,gamper2008compressed}. Sparsity can also be exploited in multiple domains, for example in both temporal and spatial domains \citep{lustig2006kt}. 

Here, we focus on Cartesian sampling, so that $\mat{F}_u$ represents lines samples of the {\bf\emph k}-space in 2D imaging. We further consider brain MRI, known to be sparse in the wavelet domain. Therefore, we will assume throughout that $\mat{\Psi}$ is an appropriately chosen wavelet transform. 

\subsection{The sparsity of longitudinal MRI scans}

\begin{figure*}

\includegraphics[width=72pt,trim=1.6cm
0.7cm 1.6cm 0.7cm,
clip=true]{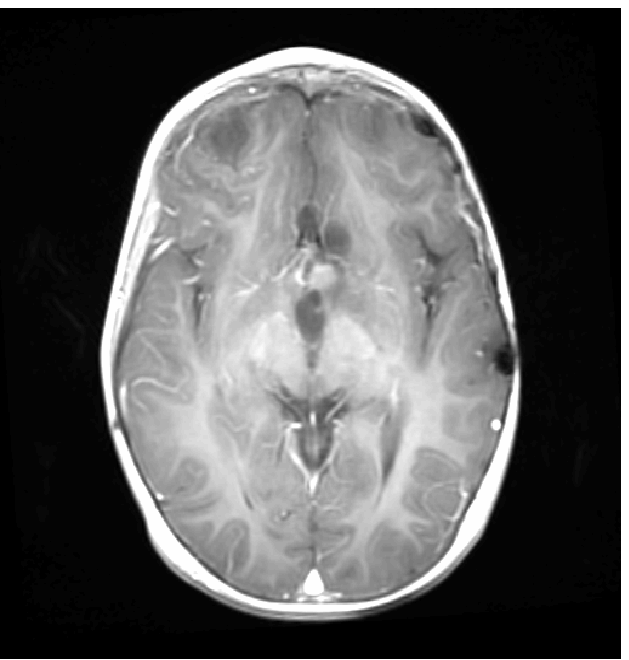}
\includegraphics[width=72pt,trim=1.6cm
0.7cm 1.6cm 0.7cm,
clip=true]{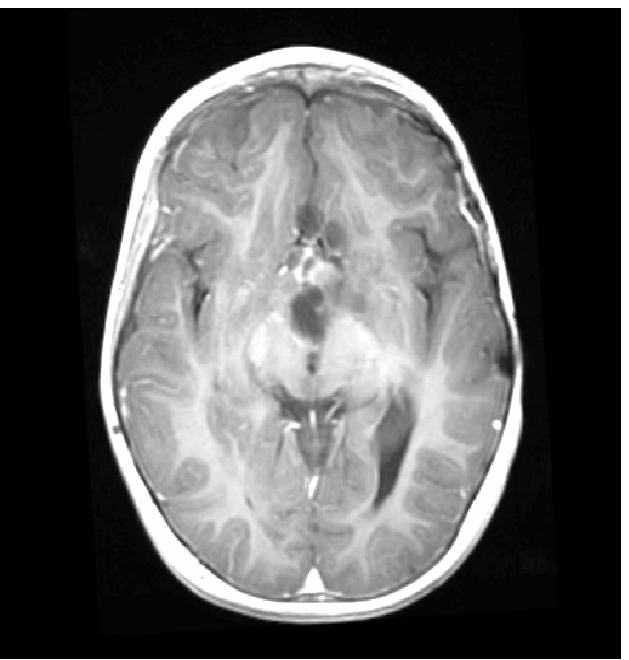}
\includegraphics[width=72pt,trim=1.6cm
0.7cm 1.6cm 0.7cm,
clip=true]{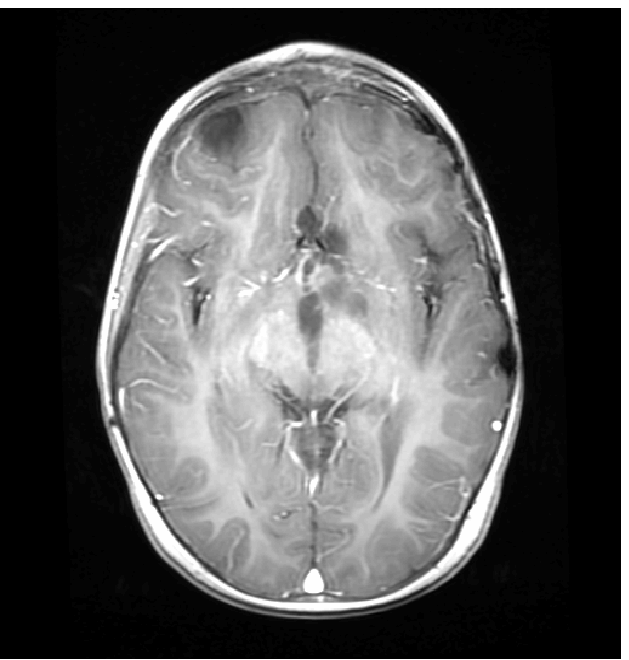}
\includegraphics[width=72pt,trim=1.6cm
0.7cm 1.6cm 0.7cm,
clip=true]{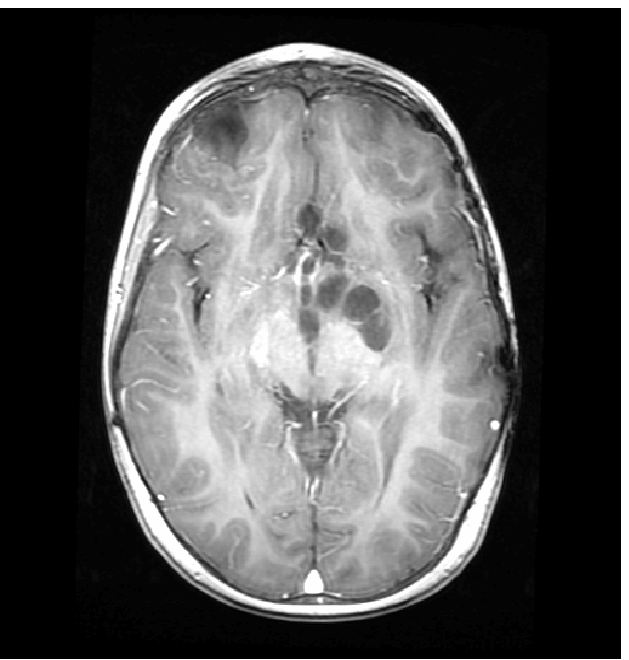}
\includegraphics[width=72pt,trim=2.35cm
0.7cm 1.65cm 0.92cm,
clip=true]{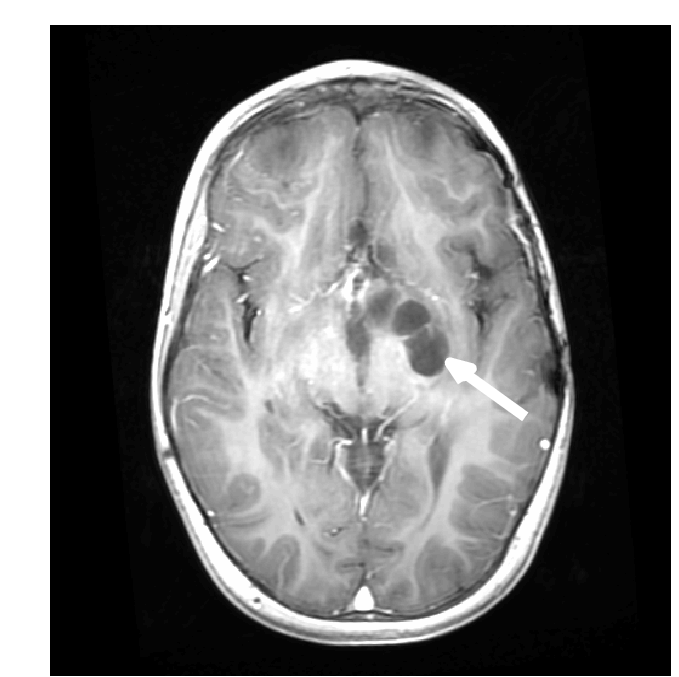}\\

\includegraphics[width=72pt,
clip=true]{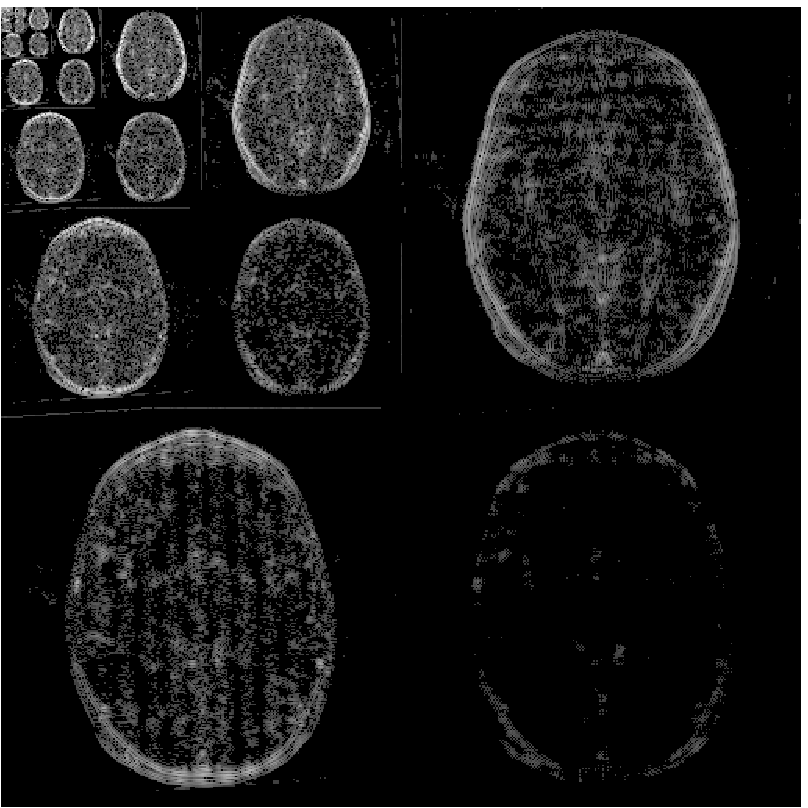}
\includegraphics[width=72pt,
clip=true]{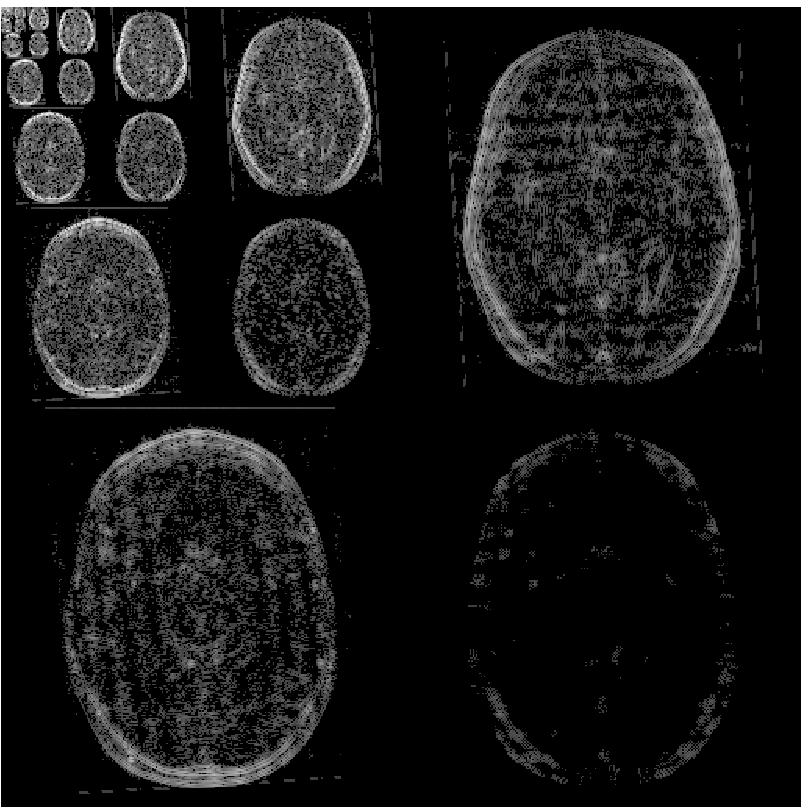}
\includegraphics[width=72pt,
clip=true]{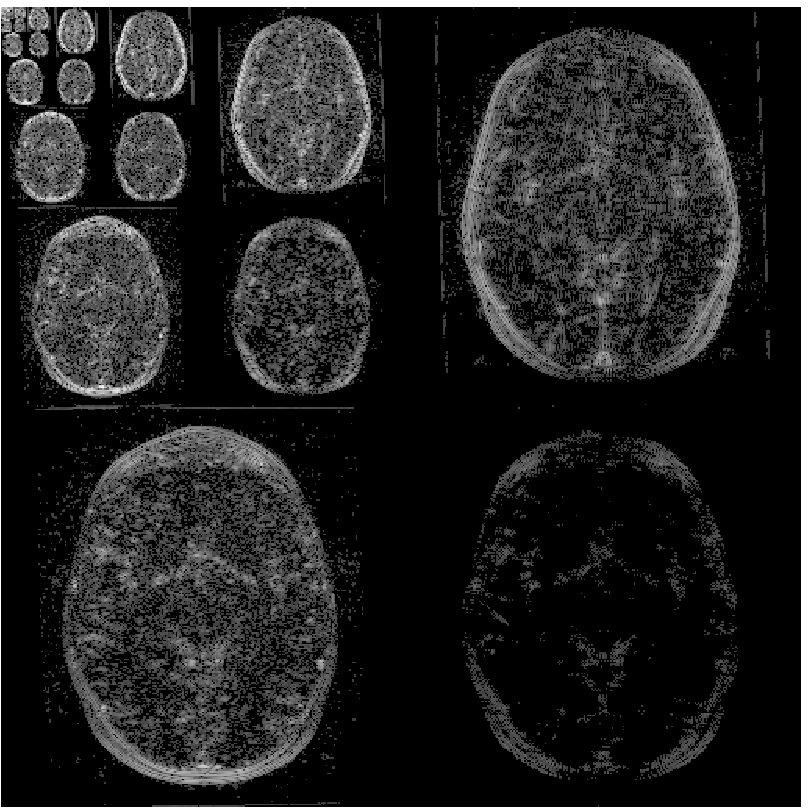}
\includegraphics[width=72pt,
clip=true]{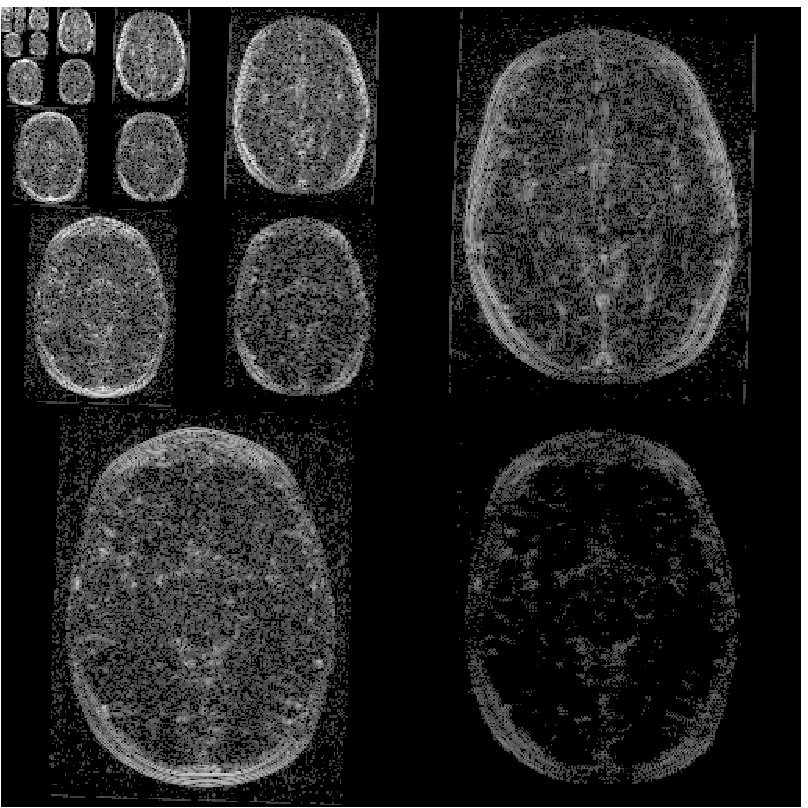}
\includegraphics[width=72pt,
clip=true]{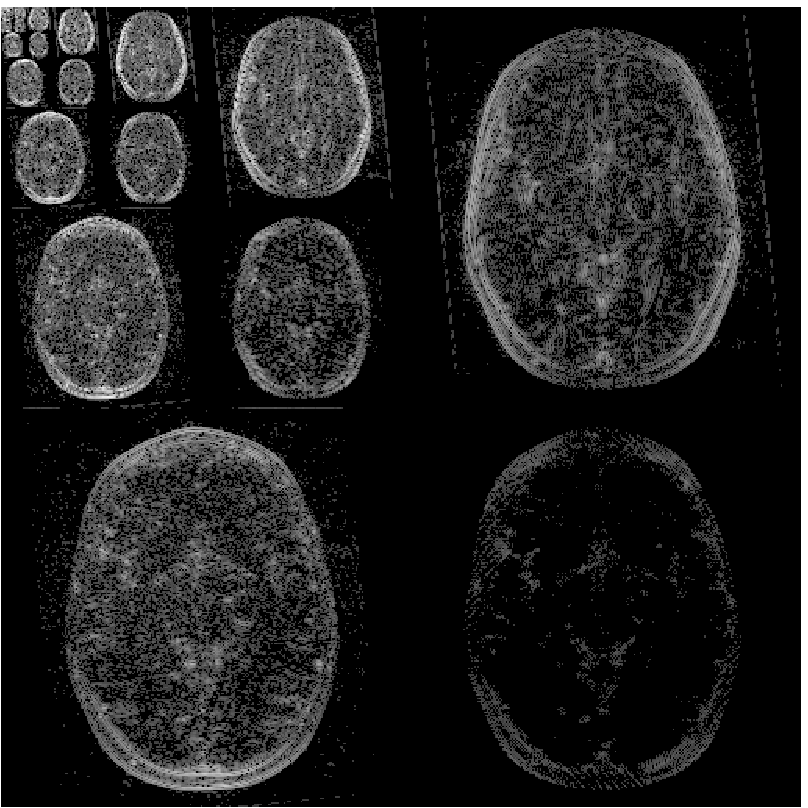}

\begin{minipage}{\linewidth}\begin{flushleft}
\vspace{2mm}
\hspace{4mm} Baseline \hspace{8mm} 5 months \hspace{5mm}  10 months \hspace{5mm} 20 months \hspace{5mm} 23 months
\end{flushleft}
\end{minipage}
\caption[Brain contrast-enhanced T1-weighted longitudinal scans of patient with Optic Pathway Gliomas (OPG)]{Brain contrast-enhanced T1-weighted longitudinal scans of patient with Optic Pathway Gliomas (OPG). The same spatial slice (top) and its representation in the wavelet domain (bottom) are shown before treatment (leftmost image) and 5, 10, 20 and 23 months after the beginning of treatment. It can be seen that despite the cyst evolving over time (marked by an arrow in the rightmost image) many image regions remain unchanged from one time point to another. Moreover, the support of the image in the wavelet domain (bottom) is preserved with no major changes over time (the minor structured artifacts in the wavelet domain are related to image registration between time points).}
\label{fig1}

\end{figure*}

In longitudinal studies, patients are scanned every several weeks or months. This scanning scheme is broadly used, for follow-up purposes and for therapy response assessment. In this setting, we could save scanning time if we are able to scan only the changes from the baseline. This type of scanning is, of course, not feasible, due to practical difficulties in designing such a sequence and the fact that prior information on the changes is mostly unavailable. 

While the scenario of longitudinal studies is fundamentally different from dynamic MRI in many aspects, we still find that similarity across time points exists in many cases. Figure~\ref{fig1} shows an example of the same axial slice taken from multiple scans acquired from a patient with Optic Pathway Glioma (OPG), demonstrating a relatively slow growing tumor pattern. The bottom row of the figure shows the representation of each time point in Daubechies-4 wavelet transform \citep{daubechies1992ten}, which is widely used as a sparse transform for brain MRI. The similarity between image slices acquired at several time points is clearly demonstrated. Moreover, the representation of the images in the wavelet domain is sparse, and the locations of the dominant wavelet coefficients (a.k.a the support of the image in the wavelet domain) are similar across the patient's time points.

Exploiting temporal similarity can be embedded in (\ref{eq1}) by an additional term, which will promote sparsity of the difference between the image to be reconstructed, $\vec{x}$, and a previously acquired image of the same patient, $\vec{x}_0$. This leads to the modified problem:
\begin{equation}
\begin{aligned}
& \underset{\vec{x}}{\text{min}}
& &  \underbrace{\|\mat{\Psi}\vec{x}\|_1}_\text{\emph{term 1}}+\lambda\underbrace{\|\vec{x}-\vec{x}_0\|_1}_\text{\emph{term 2}} \ \ \text{s.t.}
&& \|\mat{F}_u\vec{x}-\vec{y}\|_2<\epsilon.
\end{aligned}
\label{eq2}
\end{equation}

Here, \emph{term 1} enforces sparsity of $\vec{x}$ in the wavelet domain, and \emph{term 2} enforces similarity of $\vec{x}$ to $\vec{x}_0$ in the image domain. The parameter $\lambda$ trades sparsity in the wavelet domain with sparsity in the temporal domain. This approach of utilizing sparsity in both spatial and temporal domains via CS is coined hereinafter TCS-MRI (Temporal Compressed Sensing).

\begin{figure*}

\includegraphics[width=90pt,trim=1.7cm
0.7cm 1.7cm 0.9cm,
clip=true]{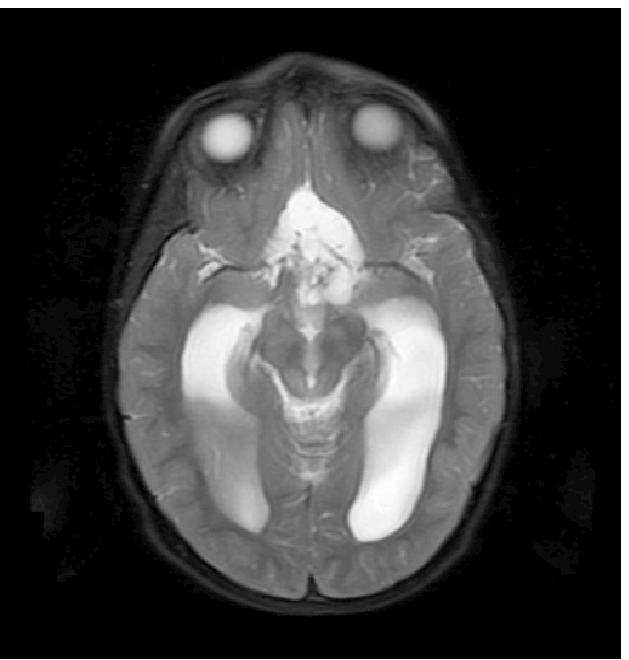}
\includegraphics[width=90pt,trim=1.7cm
0.7cm 1.7cm 0.9cm,
clip=true]{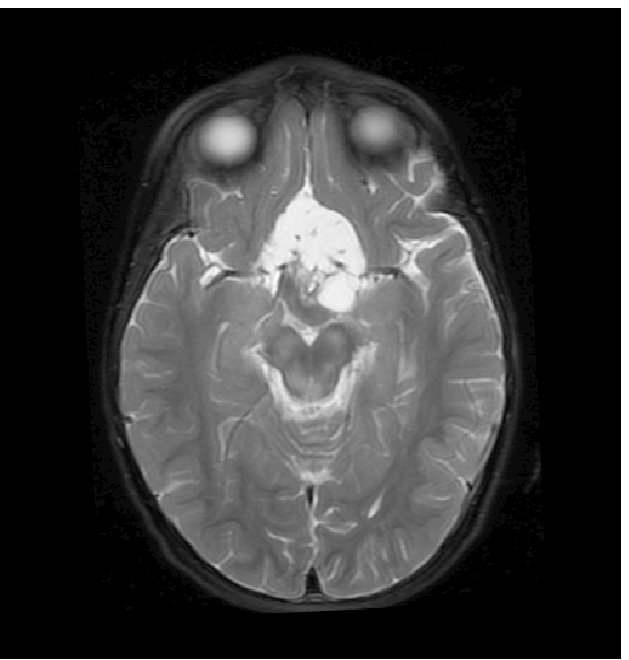}
\hspace{5mm}
\includegraphics[width=90pt,trim=1.7cm
0.1cm 1.7cm 1.5cm,
clip=true]{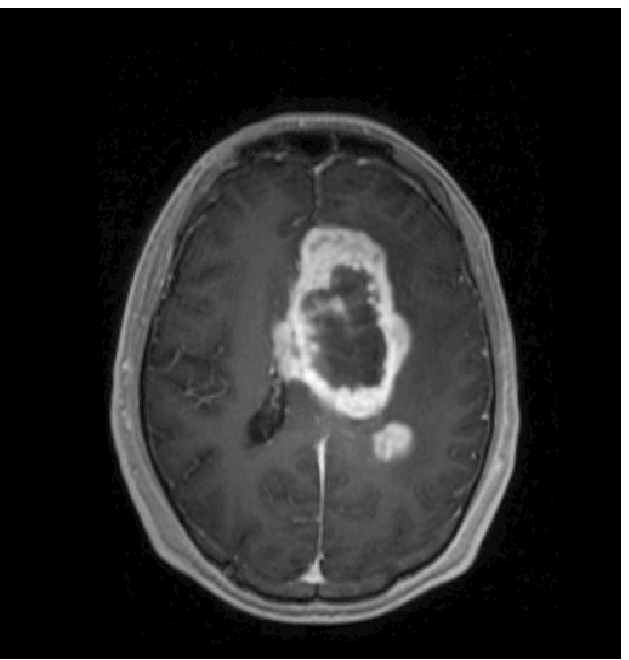}
\includegraphics[width=90pt,trim=1.7cm
0.1cm 1.7cm 1.5cm,
clip=true]{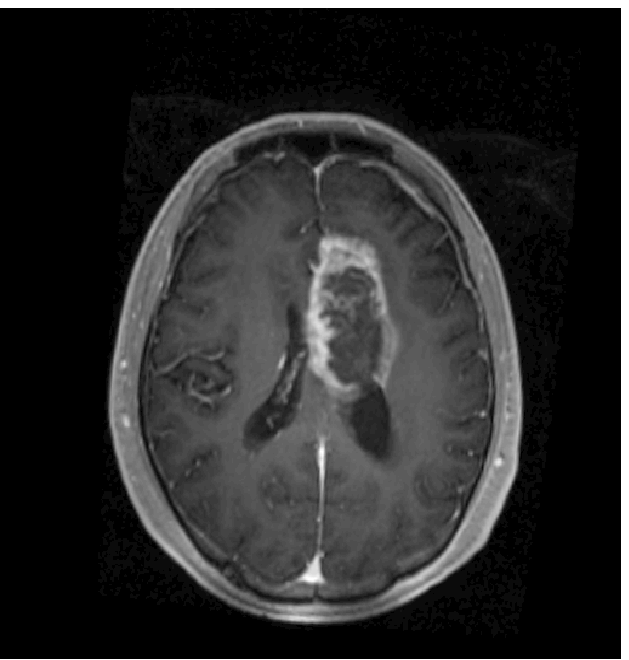}\\
\includegraphics[width=90pt,
clip=true]{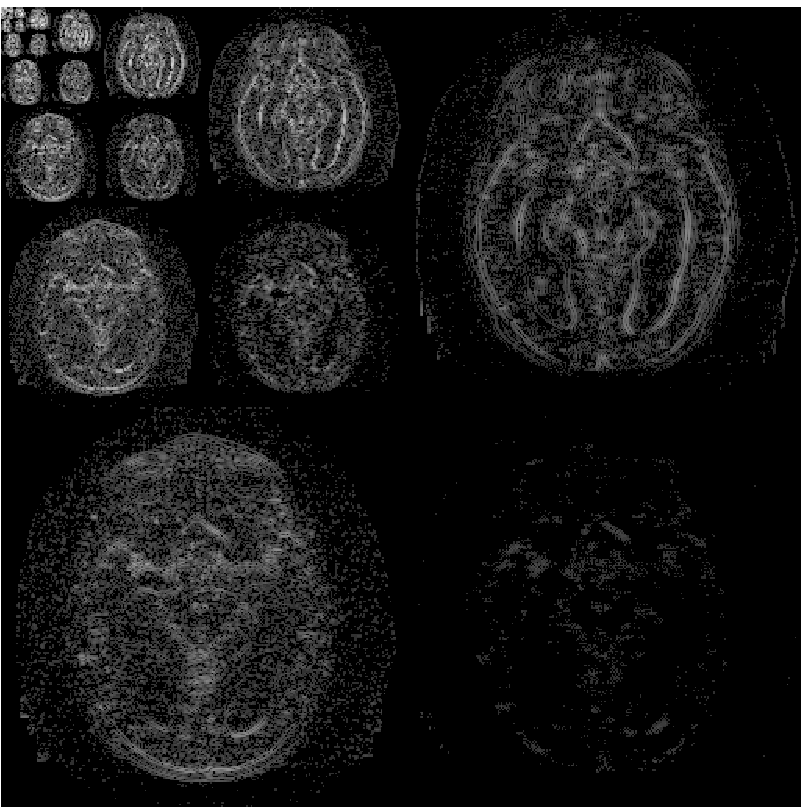}
\includegraphics[width=90pt,
clip=true]{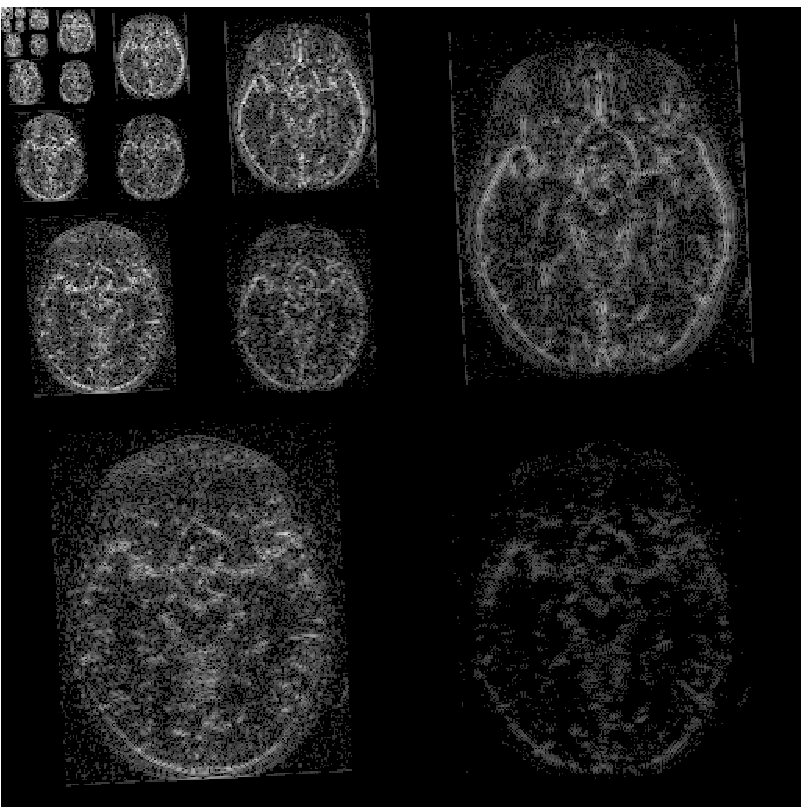}
\hspace{5mm}
\includegraphics[width=90pt,
clip=true]{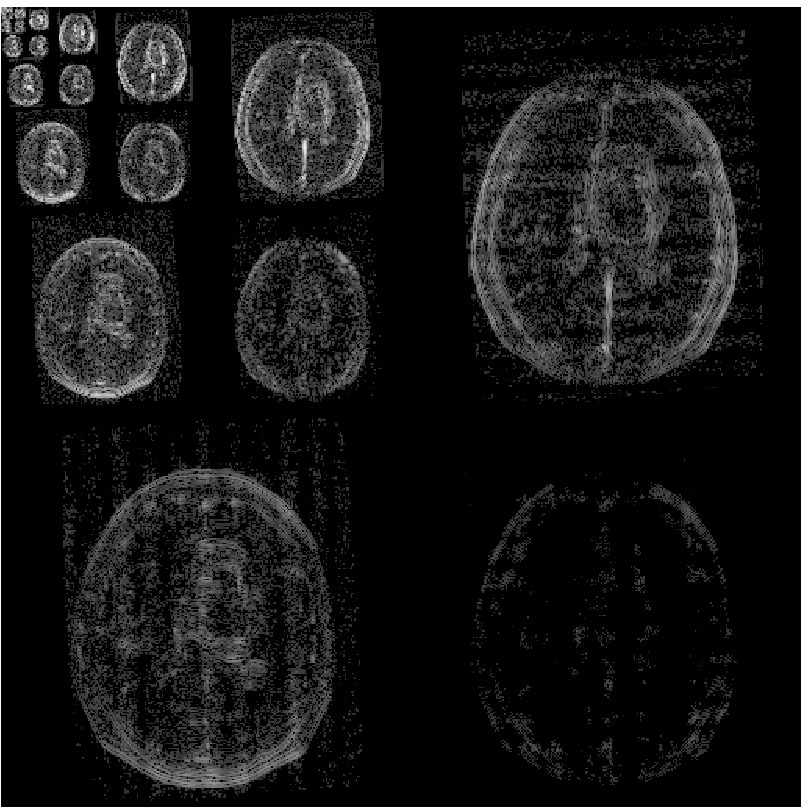}
\includegraphics[width=90pt,
clip=true]{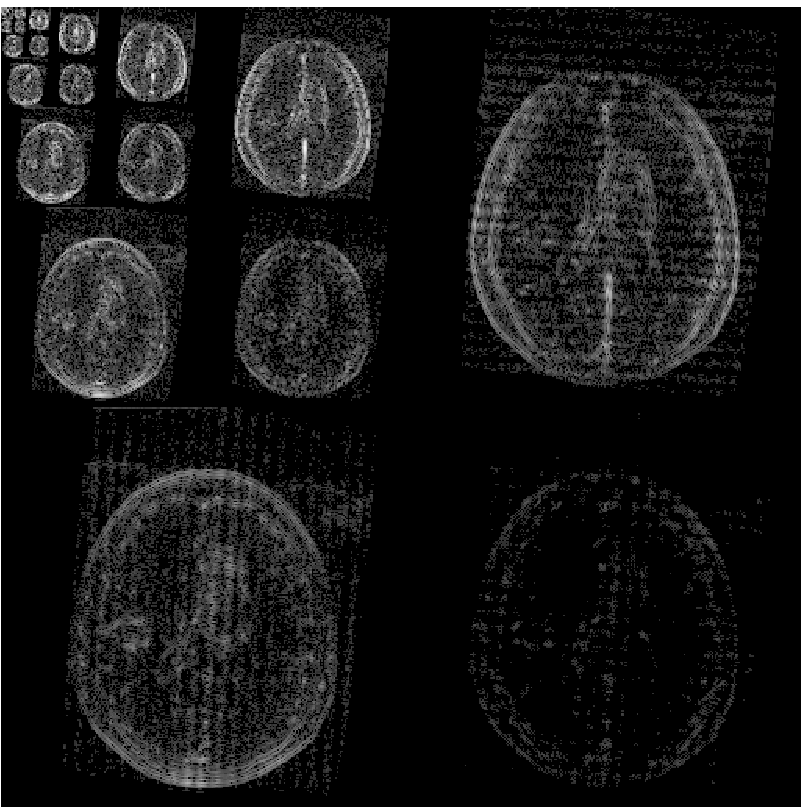}

\begin{minipage}{\linewidth}\begin{flushleft}
\vspace{1mm}
\hspace{7mm} Baseline \hspace{13mm} 4 months \hspace{14mm}  \hspace{7mm} Baseline \hspace{14mm} 3 months\\
\hspace{20mm} Hydrocephalus \hspace{45mm} GBM
\end{flushleft}
\end{minipage}
\caption[Consecutive scans of a patients with
Hydrocephalus and with Glioblastoma  multiforme (GBM)]{Left: Brain T2-weighted consecutive scans of a patient with
Hydrocephalus. Scans were acquired before and after treatment.  Right: contrast-enhanced
T1-weighted consecutive scans of a a patient with Glioblastoma  multiforme (GBM) acquired
during the treatment period. In these cases, there is substantial difference over time between the consecutive scans of the patients, in both the image (top) and the wavelet (bottom) domains.}
\label{fig2}

\end{figure*}

Note that \emph{term 2} is sparse if there are no major changes between $\vec{x}$ and $\vec{x}_0$, both images have similar grey-level intensities and they are spatially matched. While these conditions meet in many application of dynamic imaging, such as prior image constrained compressed sensing (PICCS) in CT \citep{chen2008prior,lauzier2012prior} and dynamic MRI \citep{jung2009k,lustig2006kt,gamper2008compressed,yip2014prior}, in longitudinal MRI none of these requirements are guaranteed. While there are solutions for miss-registration and variable grey level intensities (see Section \ref{discussion}), the temporal similarity in longitudinal MRI is a-priori unknown. Although longitudinal MRI may exhibit temporal similarity \citep{samsonov_ismrm}, we have to take into account that in many cases the follow-up scan may exhibit substantial changes with respect to the baseline scan. Such cases may occur, for example, if a surgical intervention was applied between the time points or if there is a major progressive or therapy response. Figure \ref{fig2} shows two representative examples.

Therefore, using (\ref{eq2}) for image reconstruction under the assumption of substantial similarity between time points might result in improper reconstruction in cases where this assumption does not hold.  To avoid this we have to carefully design a sampling and reconstruction  mechanism that  will be adjusted to match the temporal similarity of the case at hand. To achieve this goal, we extend TCS-MRI using weighted reconstruction and adaptive sampling.

\subsection{Adaptive sampling and Weighted reconstruction}
Many extensions have been proposed to improve the performance of the sampling and reconstruction phases. One of these extensions consists of embedding prior knowledge for better recovery by weighted reconstruction, which has been examined in many image reconstruction problems \citep{fessler1992regularized,hero1999minimax,haldar2008anatomically} and also in the context of CS \citep{vaswani2010modified,khajehnejad2009weighted,friedlander2012recovering}. Weighted reconstruction can be mathematically embedded into the basic CS equation (\ref{eq1}) by adding a diagonal weighting matrix, $\mat{W}=\text{diag}([w_1,w_2,...,w_N])$, to embed prior knowledge on the support, as proposed by Candes et al. \citep{candes2008enhancing}: 

\begin{equation}
\begin{aligned}
& \underset{\vec{x}}{\text{min}}
& & \|\mat{W}\mat{\Psi}\vec{x}\|_1 \ \ \text{s.t.}
&& \|\mat{F}_u\vec{x}-\vec{y}\|_2<\epsilon
\end{aligned}
\label{eq3}
\end{equation}

\noindent where $w_i$ represents the probability that $i \notin T$, and $T$ is the support of $\vec{x}$ in the sparse transform domain. It has been shown that weighted reconstruction outperforms traditional CS-MRI for dynamic MRI \citep{vaswani2010modified,zonoobi2012}, where the support of the previous time frame is used as an estimate for the support of the current frame. 

Prior knowledge can also be used to optimize the way that data is acquired  \citep{zientara1994dynamically,nagle1999multiple,gao2000optimal}. However, since in longitudinal MRI the similarity between scans is not guaranteed, an adaptive mechanism is needed to determine the level of similarity between scans. Adaptive sampling proposes a selective sample selection, where the choice of the next samples  depends on the previously gathered information. It has been proposed to improve signal's support detection from low number of samples or under a constraint on the total sensing effort \citep{haupt2009adaptive,haupt2011distilled,wei2013}. The rational behind this approach is that the estimation error one can get by using clever sampling based on previously acquired data, is generally lower than that achievable by a non-adaptive scheme.  While Arias-Castro et al. \citep{arias2013fundamental} have shown that the validity of this claim is limited in general, it is proven to be valid for many cases. 

In MRI, Seeger et al. \citep{seeger2010optimization} employ this concept to optimize {\bf\emph k}-space trajectories, by formulating the optimization problem as a Bayesian experimental design problem. They use the posterior uncertainty as the criterion for selecting the next trajectory at each round. Ravishankar and Bresler \citep{ravishankar2011adaptive} propose an adaptive scheme that relies on training image scans to optimize the sampling pattern. Their criterion takes into account training data and the reconstruction strategy. 


In the application of CS for longitudinal studies we use weighted reconstruction and adaptive sampling  in a ``patietnt-specific" way:  {\bf\emph k}-space trajectories will be optimized based on the past scan of the patient currently being scanned and reconstruction is improved via weighted reconstruction. In addition, this approach will iteratively detect cases in which the assumption of temporal similarity does not hold, and will update the sampling and reconstruction processes accordingly. 
\subsection{Adaptive-Weighted CS for Longitudinal MRI}
Embedding weighted reconstruction into the longitudinal MRI results in the following minimization problem: 
\begin{equation}
\begin{aligned}
& \underset{\vec{x}}{\text{min}}
& & \underbrace{\|\mat{W}_1\mat{\Psi}\vec{x}\|_1}_\text{\emph{term 1}}+\lambda\underbrace{\|\mat{W}_2(\vec{x}-\vec{x}_0)\|_1}_\text{\emph{term 2}} \ \  \text{s.t.}
&& \|\mat{F}_u\vec{x}-\vec{y}\|_2<\epsilon
\end{aligned}
\label{eq4}
\end{equation}
\vspace{2mm}
\noindent where $\mat{W}_k$ is a diagonal matrix, $\mat{W}_k=\text{diag}([w_k^1,w_k^2,...,w_k^N])$ and $w_k^i$ controls the weight given to each element in the support of {\emph{term 1}} or {\emph{term 2}}. Adding $\mat{W}_1$ to \emph{term 1} relaxes the demand for sparsity on the elements in the support of the image in its sparse transform domain. As a result, sparsity in the wavelet domain is strongly enforced on elements outside of the support. Adding $\mat{W}_2$ to \emph{term 2} controls the demand for similarity between $\vec{x}$ and $\vec{x}_0$. As a result, sparsity is enforced only in image regions where $\vec{x}$ and $\vec{x}_0$ are similar.

The solution of problem (\ref{eq4}) can be obtained via extending one of the well-known approaches for the classical CS problem \citep{donoho2008fast,becker2011nesta}. In our experiments, we extended the fast iterative shrinkage-thresholding algorithm (FISTA) \citep{beck2009fast} to solve the unconstrained problem in so-called Lagrangian form:
\begin{equation}
\begin{aligned}
& \underset{\vec{x}}{\text{min}}
&  \|\mat{F}_u\vec{x}-\vec{y}\|_2^2+\lambda_1\underbrace{\|\mat{W}_1\mat{\Psi}\vec{x}\|_1}_\text{\emph{term 1}}+\lambda_2\underbrace{\|\mat{W}_2(\vec{x}-\vec{x}_0)\|_1}_\text{\emph{term 2}}
\end{aligned}
\label{eq5}
\end{equation}
The values of $\lambda_1$ and $\lambda_2$ can be selected appropriately such that the solution of (\ref{eq5}) is exactly as (\ref{eq4}), for a given $\lambda$. These values control the trade-off between enforcing sparsity in the wavelet and temporal domains. 
Detailed implementation of our FISTA-based approach can be found in Appendix A.

%
When determining the values of $\mat{W}_1$ and $\mat{W}_2$ we would like to avoid utilizing the prior scan in the reconstruction process if the assumption of similarity between consecutive scans does not hold. 
More specifically, we design $\mat{W}_1$ and $\mat{W}_2$ to achieve the following goals:
\begin{enumerate}
  \item Convergence to CS-MRI (\ref{eq1}) if the assumption of similarity between consecutive scans is not valid (i.e.: $w_1^i \rightarrow 1$ and $w_2^i \rightarrow 0$ if $\vec{{x}}$ and $\vec{x}_0$ are significantly different in both image and wavelet domains).
  \item Relaxing the demand for sparsity of $\mat{\Psi}\vec{x}$ in regions where $\mat{\Psi}\vec{x}_0$ is not sparse and the similarity assumption between $\mat{\Psi}\vec{x}$ and $\mat{\Psi}\vec{x}_0$ is valid. (i.e.: $w_1^i \rightarrow 0$ as $[|\mat{\Psi}\vec{{x}_0}|]_i$ grows and images are similar in the transform domain)
  \item  Relaxing the demand for sparsity of $(\vec{x}-\vec{x}_0)$ in regions where the similarity assumption between $\vec{x}$ and $\vec{x}_0$ is not valid (i.e.: $w_2^i \rightarrow 0$ as $[|\vec{{x}}-\vec{x}_0|]_i$ grows). 
\end{enumerate} 
To obtain the goals above, we first sample $N_k$ {\bf\emph k}-space samples randomly and reconstruct $\vec{\hat{x}}$, which is the estimation of $\vec{x}$, by solving (\ref{eq5}) with $\mat{W}_1=\mat{I}$ and $\mat{W}_2=\mat{0}$. We then sample additional $N_k$ samples and solve (\ref{eq5}) where the elements of the matrices are chosen as follows:
\begin{equation}
 w_1^i=\begin{cases}
    1, &  \frac{[|\mat{\Psi}(\vec{\hat{x}}-\vec{x}_0)|]_i}{1+[|\mat{\Psi}(\vec{\hat{x}}-\vec{x}_0)|]_i}>\epsilon_1 \\
   \frac{1}{1+[|\mat{\Psi}\vec{x}_0|]_i}, & otherwise
  \end{cases}
\label{eq6}
\end{equation}
\begin{equation}
w_2^i=\frac{1}{1+[|\vec{\hat{x}}-\vec{x}_0|]_i}
\label{eq7}
\end{equation}
\noindent where $[\cdot]_i$ denotes the $i$th element of the vector in brackets and $\epsilon_1$ is a threshold for defining similarity in the sparse transform domain. This process is repeated until a sufficient number of samples has been obtained for adequate recovery.  This iterative approach allows exploitation of temporal similarity, when it exists, and prevents degradation of image quality if the consecutive scans are significantly different. 


Incorporating adaptive sampling into the longitudinal MRI problem is obtained via adaptive design of the $N_k$ sampling locations at each iteration. It is well known that reconstruction results highly depend with sampling trajectories in {\bf\emph k}-space domain. For instance, random sampling is one of the requirement of CS and one should be undersampling less near the {\bf\emph k}-space origin and more in the periphery of
{\bf\emph k}-space \citep{tsai2000reduced,lustig2007sparse}. A common sampling scheme is variable density random undersampling (VDS), which can be implemented by choosing samples randomly with sampling density scaling according to a power of distance from the origin. 

For simplicity, we assume that 2D Cartesian sampling is used, and the $N_k$ sampling locations consist of rows sampling in the {\bf\emph k}-space domain. To utilize VDS in 2D Cartesian sampling we define the discrete polynomial distribution as: 
\begin{equation}
f_{VD}(k_y)=\frac{(1-\frac{2}{n}|k_y|)^p}{\sum\limits_{k_y}(1-\frac{2}{n}|k_y|)^p}
\label{eq8}
\end{equation}
\noindent where $-\frac{n}{2}<k_y\le \frac{n}{2}$ denotes the {\bf\emph k}-space coefficients in the phase encoding direction and $p$ is the power distance from the origin. 


VDS is used as the probability density function (pdf) for random sampling in many cases where the real distribution of the data is a-priori unknown. However, in longitudinal studies we may rely on the reference scan data distribution, if scans are similar. Inspired by Chen et al. \citep{chen2014completing}, our adaptive approach will converge to random sampling according to the reference scan data pdf if it is similar to the follow-up scan, and to polynomial pdf otherwise. Therefore, samples in our approach are taken randomly using the following pdf:
\begin{equation}
f_S(k_y)=\gamma f_{B}(k_y)+(1-\gamma)f_{VD}(k_y)
\label{eq9}
\end{equation}
\noindent where $f_{B}$ is the pdf of the baseline's phase encode lines' energy, defined as:
\begin{equation}
f_{B}(k_y)=\frac{g_{B}(k_y)}{\sum\limits_{k_y}g_{B}(k_y)}\ \  \ \ \ ,  \ \ \ \ \ \ \ 
g_{B}(k_y)=\sum\limits_{i\in k_y}{[|\mat{F}\vec{{x_0}}|]_i} 
\label{eq10}
\end{equation}
\noindent $\mat{F}$ indicates the $N \times N$ Fourier matrix, $[\cdot]_i$ denotes the $i$th element of the vector in brackets and $\gamma$ is the fidelity we give to the similarity between the current and the previous scan. Since $\mat{W}_2$ can serve as a good approximation for this similarity, $\gamma$ is computed as its mean over the main diagonal: $\gamma=\frac{1}{N}\sum_{i=1}^{N}{w_2^i}$.
\begin{figure*}
\includegraphics[width=400pt]{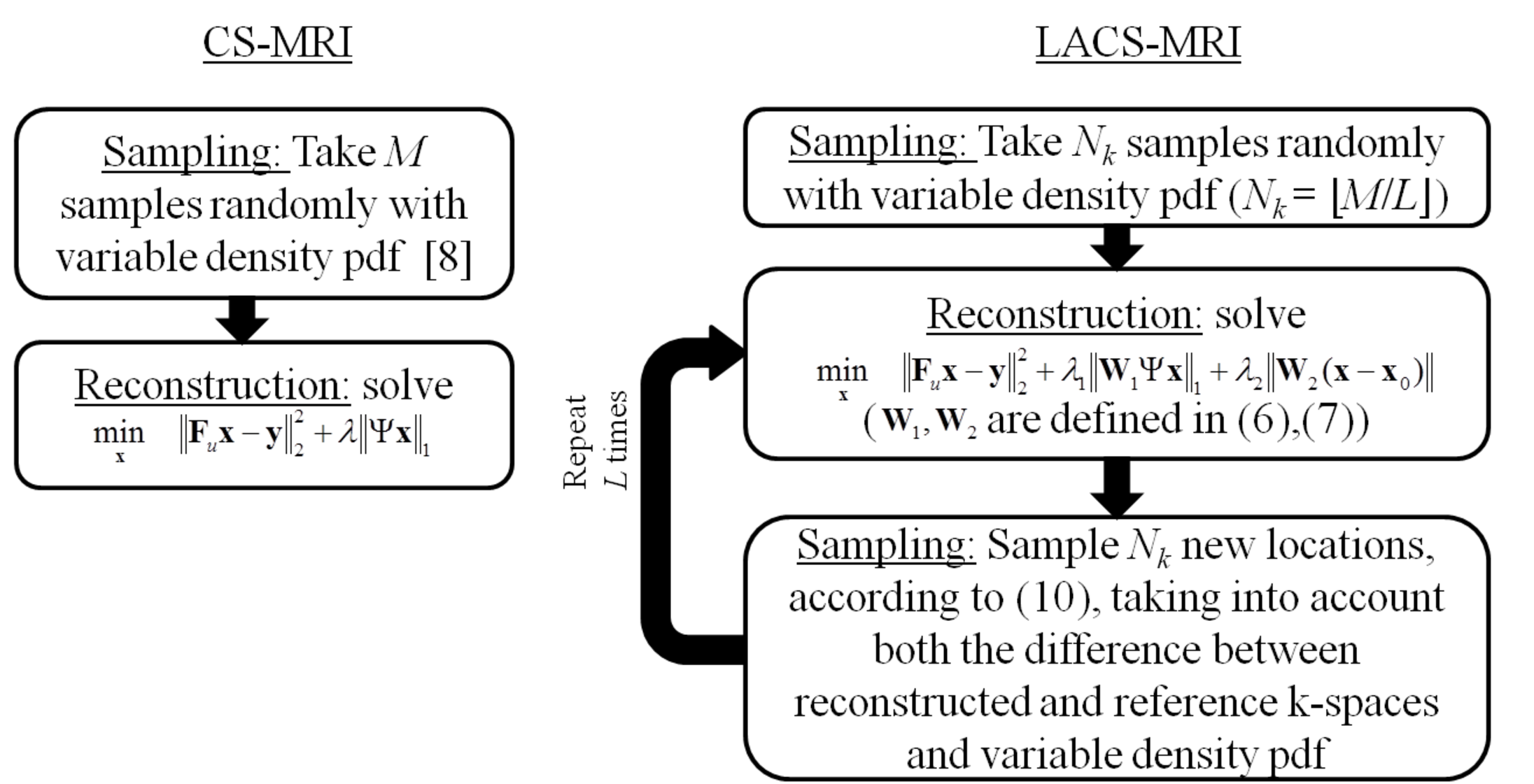}
\caption[The main steps in the CS-MRI approach and our proposed adaptive approach (LACS-MRI)]{High-level diagram showing the main steps in the CS-MRI approach (left) and our proposed adaptive approach (right). Both approaches acquire $M$ {\bf\emph k}-space samples. However, in LACS-MRI only a small portion of the samples is taken from a pure variable density pdf, in the first iteration. The rest of the samples are taken intertively, taking into account both the difference between the {\bf\emph k}-spaces of the reconstructed image, $\vec{x}$, and the previous image in the time series, $\vec{x}_0$ and the variable density pdf.}
\label{fig3}
\end{figure*}

A diagram presenting the major differences between CS-MRI, TCS-MRI and the proposed method is shown in Fig. \ref{fig3}. The details of the proposed approach are summarized in Algorithm 1. This scheme for longitudinal adaptive-weighted compressed sensing is coined hereinafter LACS-MRI (Longitudinal Adaptive Compressed Sensing MRI).


\newlength\myindent
\setlength\myindent{10em}
\newcommand\bindent{%
  \begingroup
  \setlength{\itemindent}{\myindent}
  \addtolength{\algorithmicindent}{\myindent}
}
\newcommand\eindent{\endgroup}
\renewcommand{\algorithmicrequire}{\textbf{Input:}}
\renewcommand{\algorithmicensure}{\textbf{Output:}}
\begin{algorithm}[H]
\caption{ Adaptive-weighted temporal CS MRI (LACS-MRI)}
\label{algo1}
 \begin{algorithmic} 
 \renewcommand{\algorithmicrequire}{\textbf{Input:}}
\renewcommand{\algorithmicensure}{\textbf{Output:}}
\REQUIRE \hspace{3mm} \\
Number of {\bf\emph k}-space samples acquired at each iteration: $N_k$;\\
Number of iterations: $L$; \\
Image from earlier time-point: $\vec{x}_0$;\\
\ENSURE Estimated image: $\mat{\hat{x}}$
 \renewcommand{\algorithmicrequire}{\textbf{Initialize:}}
 \REQUIRE \hspace{3mm} \\
$\vec{y}=\vec{0}$; $\mat{F}_u=\mat{0}$; $S=\varnothing$; $\mat{W}_1=\mat{I}$; $\mat{W}_2=\mat{0}$;\\
Randomly define $N_k$ {\bf\emph k}-space sampling locations: $S^{(1)}=\{s_1,s_2,...,s_{N_k}\}$ according to (\ref{eq8})\\
\renewcommand{\algorithmicrequire}{\textbf{Sampling and reconstruction:}}
 \REQUIRE \hspace{3mm} \\
\FOR{$l=1$ to $L$}
\STATE \ \ \ $S \leftarrow S \cup S^{(l)}$
\STATE \ \ \ Define undersample operator: $
\mat{F}^{(l)}_u(i,j)=\begin{cases}
    \mat{F}(i,j), & i \in S\\
    0, & i \notin S
  \end{cases}$
\STATE \ \ \ {\bf Sample: }
$\vec{y}^{(l)}=\mat{F}^{(l)}_u\vec{x}$
\STATE \ \ \ {\bf Update: } 
$\vec{y}=\vec{y}+\vec{y}^{(l)}$; $\mat{F}_u=\mat{F}_u+\mat{F}^{(l)}_u$
\STATE \ \ \ {\bf Weighted reconstruction:} \\
\ \ \  \ \   $\vec{\hat{x}}=\underset{\vec{x}}{\text{min}}
  \|\mat{F}_u\vec{x}-\vec{y}\|_2^2+\lambda_1\|\mat{W}_1\mat{\Psi}\vec{x}\|_1+\lambda_2\|\mat{W}_2(\vec{x}-\vec{x}_0)\|_1$
\STATE \ \ \ {\bf Adaptive sampling: }\\
\ \ \ \    Update $\mat{W}_1$ and $\mat{W}_2$ according to (\ref{eq6}) and (\ref{eq7}) \\
\ \ \ \    Use pdf in (\ref{eq9}) to randomly define $N_k$ new sampling locations, $S^{(l+1)}$\\
\ENDFOR
\end{algorithmic}
\end{algorithm}

\section{Experimental results}
\label{results}
\subsection{Experimental settings}
Our experiments consist of repeated scans of pomelo and patients with OPG and GBM. The patients with OPG were scanned with a GE Signa 1.5T HDx scanner and the pomelo and GBM experiments were performed on a GE Signa 3T HDXT scanner. In our experiments, partial {\bf\emph k}-space acquisition was obtained by down-sampling, in a software environment, a fully sampled {\bf\emph k}-space. All CS reconstructions were implemented in Matlab (The MathWorks, Natick, MA). We compare our approach (LACS-MRI) with two non-adaptive schemes, CS-MRI and temporal CS-MRI (TCS MRI). CS-MRI is described in \citep{lustig2007sparse} and TCS-MRI takes into account the knowledge of the prior scan in the time sequence, by solving (\ref{eq2}) for image reconstruction.

In our experiments we used the Daubechies 4 wavelet transform. Reconstructions based on non-adaptive sampling (CS-MRI and TCS-MRI) were obtained with variable density undersampling, where samples were taken randomly according to the pdf in (\ref{eq8}) with $p=4$. In 2D Cartesian sampling 5\% of the phase encode lines, located in its center were acquired in full. In 3D Cartesian sampling 1\% of the phase encode plane was acquired in full. For LACS-MRI we used the same variable density random pattern for the first iteration, while subsequent samples were taken according to Algorithm 1.  

LACS-MRI and TCS-MRI $\ell_1$-minimization problems were solved in their Lagrangian form using the FISTA algorithm, described in Appendix A. CS-MRI was tested both with our FISTA-based implementation of solving (\ref{eq1}) and with the non-linear conjugate sub-gradient algorithm as described in \citep{lustig2007sparse}, which adds a Total Variation (TV) \citep{tsaig2006extensions} penalty to (\ref{eq1}), where the best results (in terms of resolution improvement) are shown. 
%
The threshold for defining similarity in the sparse transform domain was set to $\epsilon_1=0.1$. In all experiments, different values of $\lambda_1,\lambda_2$ in the range of $[0.001,0.9]$ were examined, and the best result is shown for each reconstruction algorithm.  

For quantitative evaluation, the Signal-to-Noise ration (SNR) is computed for each reconstruction result, as: $SNR=10log_{10}{(\sigma_{\vec{x}}/{V_s})}$, where $\sigma_{\vec{x}}$ denotes the variance of the values in $\vec{x}$ and $V_s$ is the Mean Square Error (MSE) between the original image, $\vec{x}$ and the reconstructed image, $\hat{\vec{x}}$. Note that in some cases clinical decisions are based on subtle changes between the baseline and the follow-up scans, that are not reflected in the SNR. Therefore, results are presented both visually and quantitatively.

\subsection{2D Cartesian Sampling of Static Pomelo}
First, we acquired T1-weighted pomelo image using a SE sequence (matrix: $256\times 256$, res=$1mm$, 35 slices with $6mm$ thickness and no gap, $TR/TE=600/8ms$, flip angle=$90^{\circ}$). Then, we injected a contrast agent into the pomelo in order to create structural changes in the pomelo  without changing its spatial orientation. We then repeated the scan with the same acquisition parameters to obtain a post-contrast T1-weighted (T1c) image. As a result, we obtained two pomelo images, spatially matched, that simulate a baseline scan and a follow-up scan that consists of changes from the baseline scan. 

The aim of the simulation is to examine the performance of utilizing sparsity in both the wavelet and temporal domains with adaptive sampling for longitudinal scans, with LACS-MRI, in the absence of external artifacts such as miss-registration errors or movements during acquisition.

We performed 2D reconstruction with LACS-MRI, CS-MRI and TCS-MRI. We set the number of samples acquired at the first iteration  for LACS-MRI to $N_k=8$ lines. Results are shown with corresponding acceleration factors of 4, 6.4, and 10.6. 

\begin{figure*}
\begin{minipage}{\linewidth}\hspace{15mm}{ \parbox{3cm}{{ \singlespacing \footnotesize CS-MRI}}
\hspace{4mm}\parbox{3.0cm}{{\singlespacing \footnotesize TCS-MRI }}\hspace{9mm}\parbox{3.8cm}{{\singlespacing \footnotesize LACS-MRI }}}
\end{minipage}
\\
\begin{minipage}{0.05\linewidth}
\vspace{0.5cm}
\begin{turn}{90}\parbox{3cm}{\hspace{10mm} $\times 10.6$  }\end{turn}\hspace{0.5mm}
\end{minipage}
\begin{minipage}{0.25\linewidth}
\includegraphics[width=82pt,trim=2.5cm
2.2cm 2.5cm 2.5cm,
clip=true]{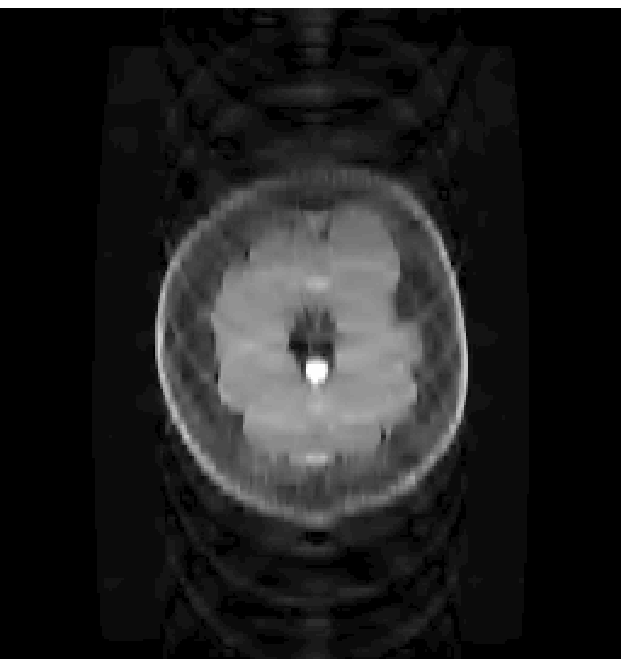}
\hspace{0.1mm}\includegraphics[width=3.3cm,angle=90,trim=0cm
5cm 0cm 5cm,
clip=true]{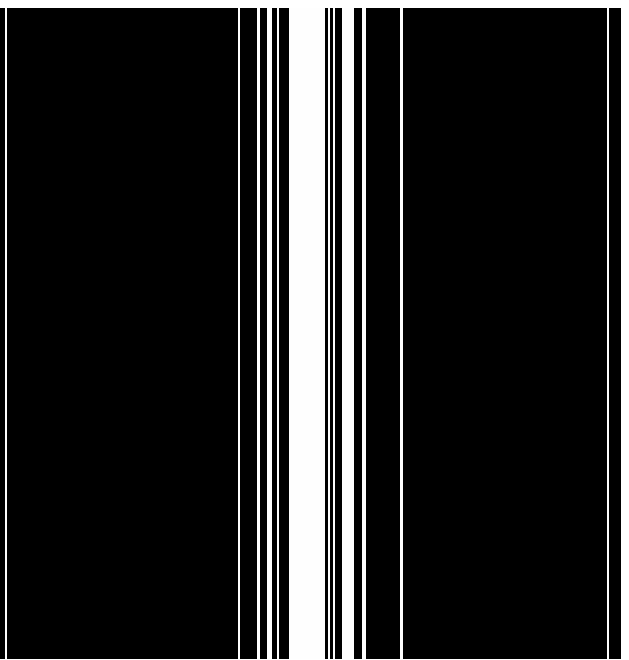}
\end{minipage}\hspace{2mm}
\begin{minipage}{0.25\linewidth}
\includegraphics[width=82pt,trim=2.5cm
2.2cm 2.5cm 2.5cm,
clip=true]{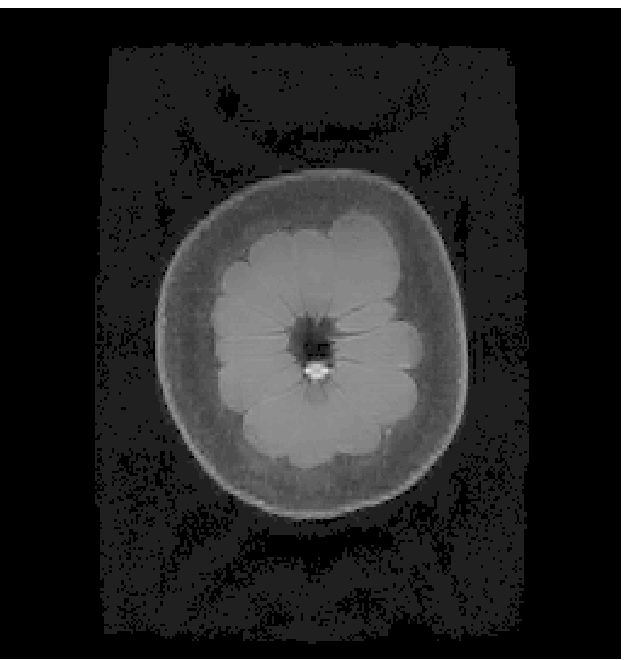}
\hspace{0.1mm}\includegraphics[width=3.3cm,angle=90,trim=0cm
5cm 0cm 5cm,
clip=true]{pomelo_non_adapative_k_space_10_fold.eps}
\end{minipage}\hspace{2mm}
\begin{minipage}{0.25\linewidth}
\includegraphics[width=82pt,trim=2.5cm
2.2cm 2.5cm 2.5cm,
clip=true]{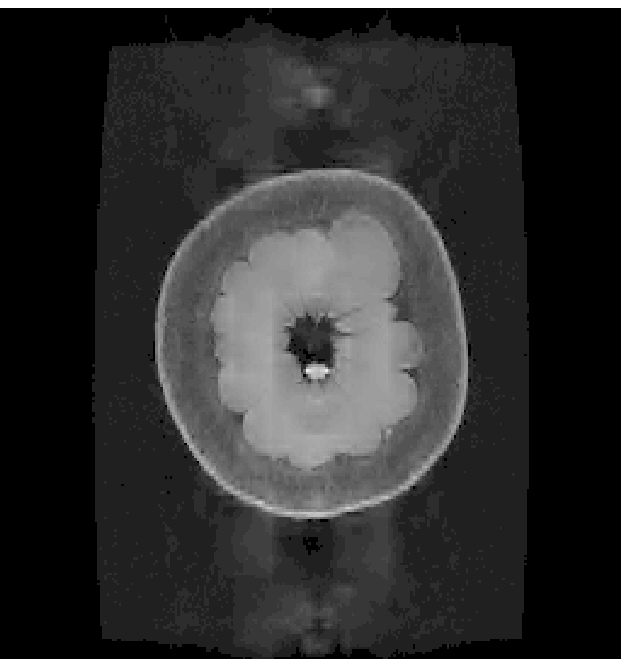}
\hspace{0.1mm}\includegraphics[width=3.3cm,angle=90,trim=0cm
5cm 0cm 5cm,
clip=true]{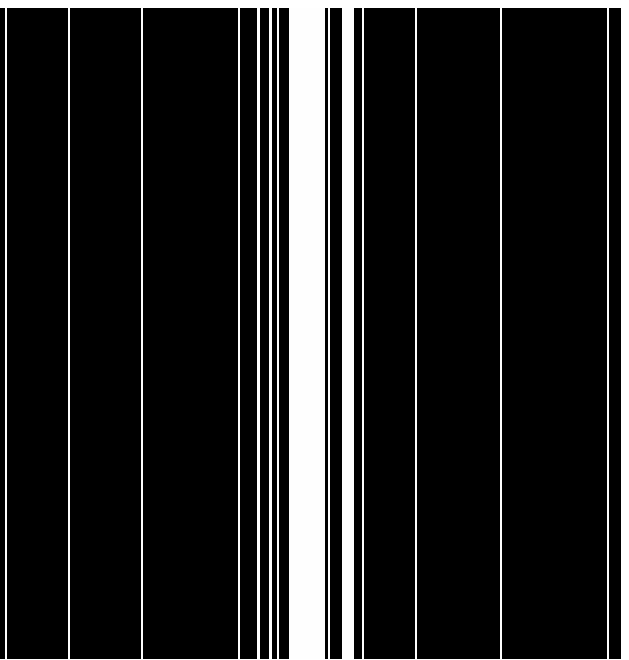}
\end{minipage}\\
\begin{minipage}{0.05\linewidth}
\vspace{0.5cm}
\begin{turn}{90}\parbox{3cm}{\hspace{10mm} $\times 6.4$  }\end{turn}\hspace{0.5mm}
\end{minipage}
\begin{minipage}{0.25\linewidth}
\includegraphics[width=82pt,trim=2.5cm
2.2cm 2.5cm 2.5cm,
clip=true]{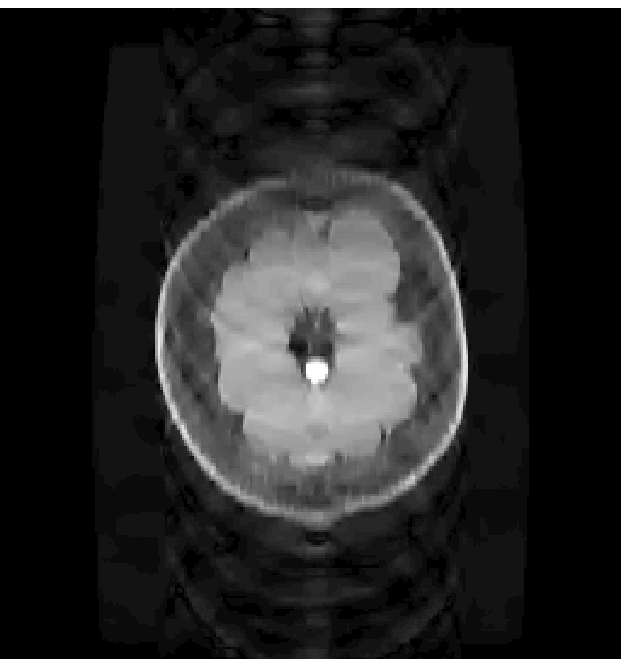}
\hspace{0.1mm}\includegraphics[width=3.3cm,angle=90,trim=0cm
5cm 0cm 5cm,
clip=true]{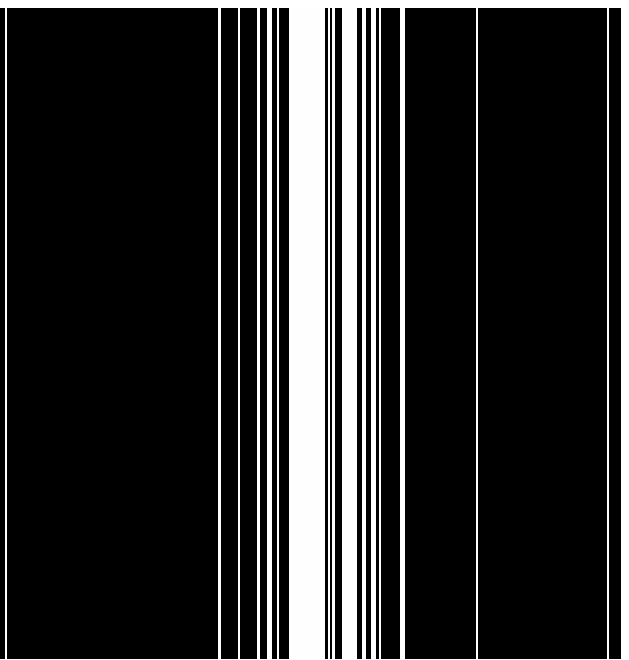}
\end{minipage}\hspace{2mm}
\begin{minipage}{0.25\linewidth}
\includegraphics[width=82pt,trim=2.5cm
2.2cm 2.5cm 2.5cm,
clip=true]{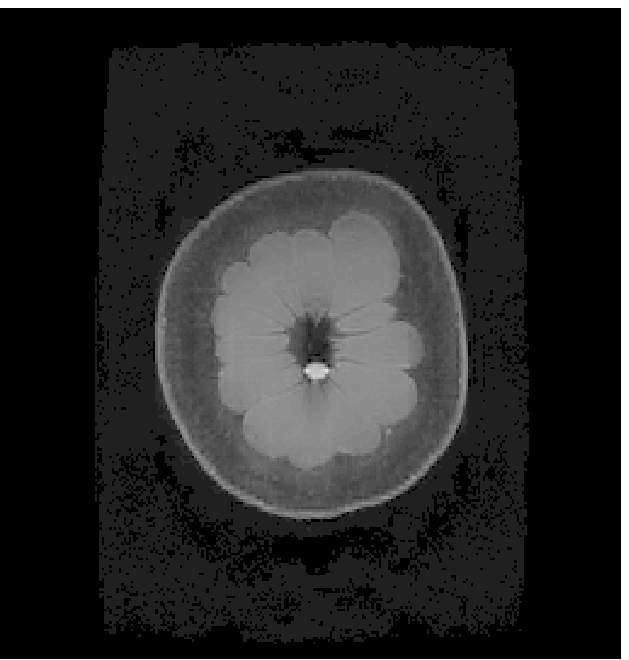}
\hspace{0.1mm}\includegraphics[width=3.3cm,angle=90,trim=0cm
5cm 0cm 5cm,
clip=true]{pomelo_non_adapative_k_space_6_fold.eps}
\end{minipage}\hspace{2mm}
\begin{minipage}{0.25\linewidth}
\includegraphics[width=82pt,trim=2.5cm
2.2cm 2.5cm 2.5cm,
clip=true]{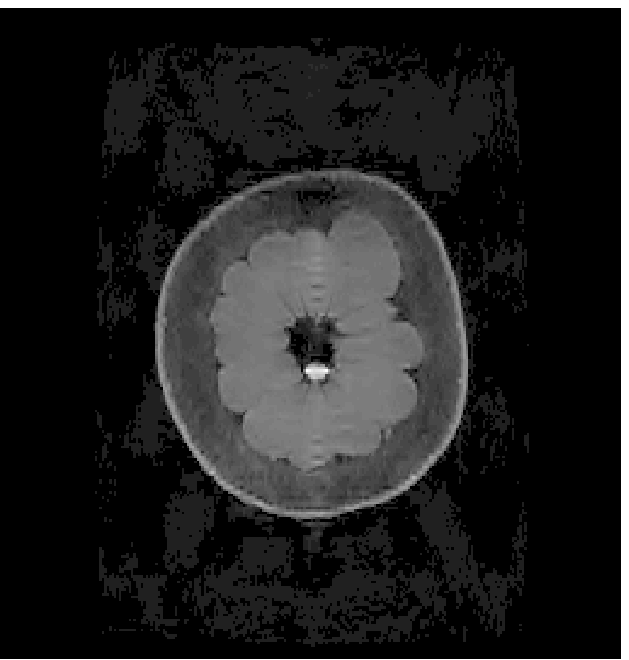}
\hspace{0.1mm}\includegraphics[width=3.3cm,angle=90,trim=0cm
5cm 0cm 5cm,
clip=true]{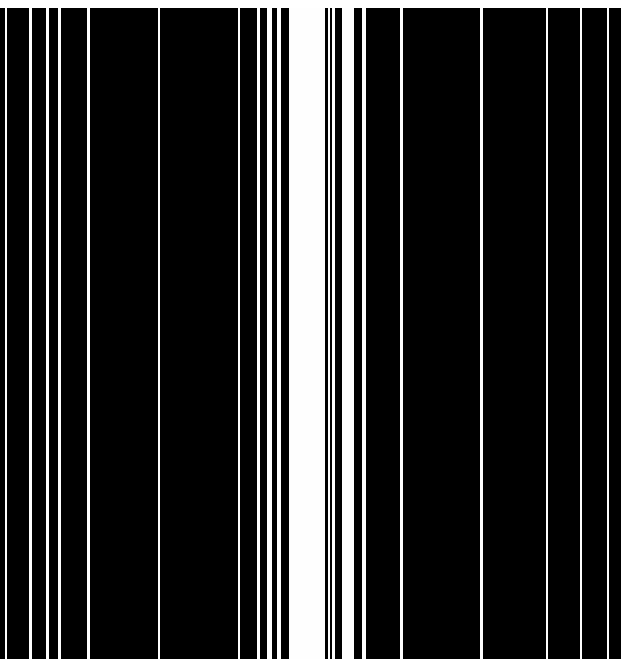}
\end{minipage}\\
\begin{minipage}{0.05\linewidth}
\vspace{0.5cm}
\begin{turn}{90}\parbox{3cm}{\hspace{12mm} $\times 4$  }\end{turn}\hspace{0.5mm}
\end{minipage}
\begin{minipage}{0.25\linewidth}
\includegraphics[width=82pt,trim=2.5cm
2.2cm 2.5cm 2.5cm,
clip=true]{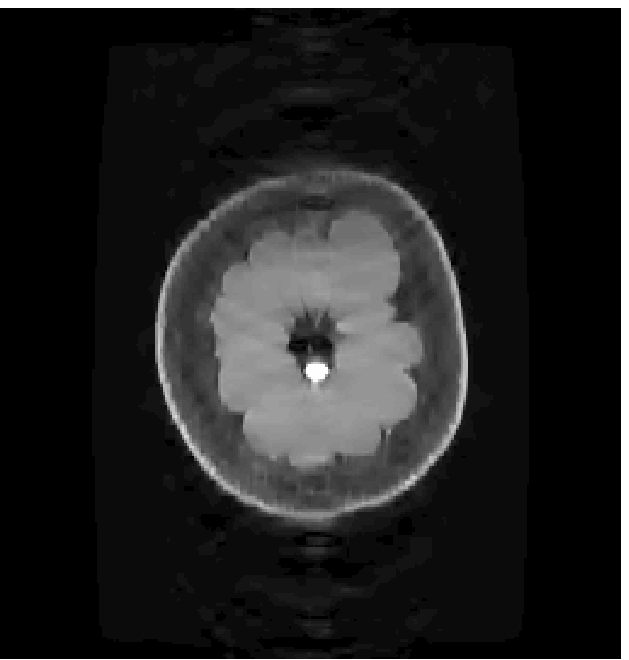}
\hspace{0.1mm}\includegraphics[width=3.3cm,angle=90,trim=0cm
5cm 0cm 5cm,
clip=true]{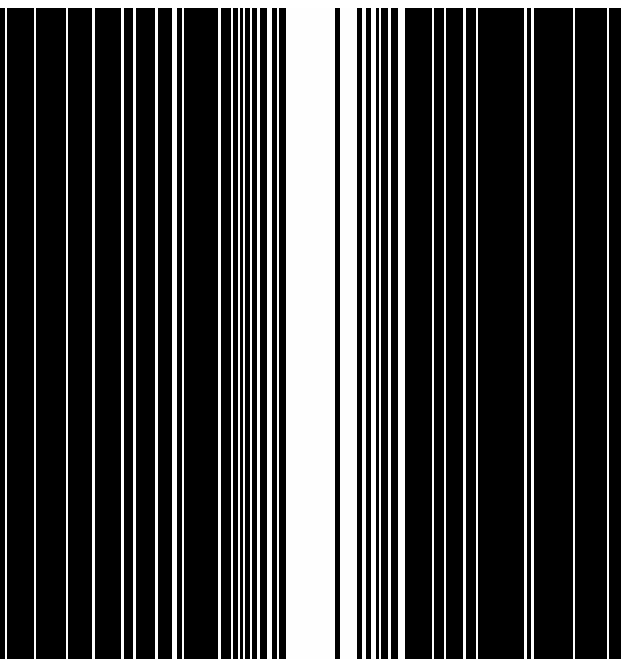}
\end{minipage}\hspace{2mm}
\begin{minipage}{0.25\linewidth}
\includegraphics[width=82pt,trim=2.5cm
2.2cm 2.5cm 2.5cm,
clip=true]{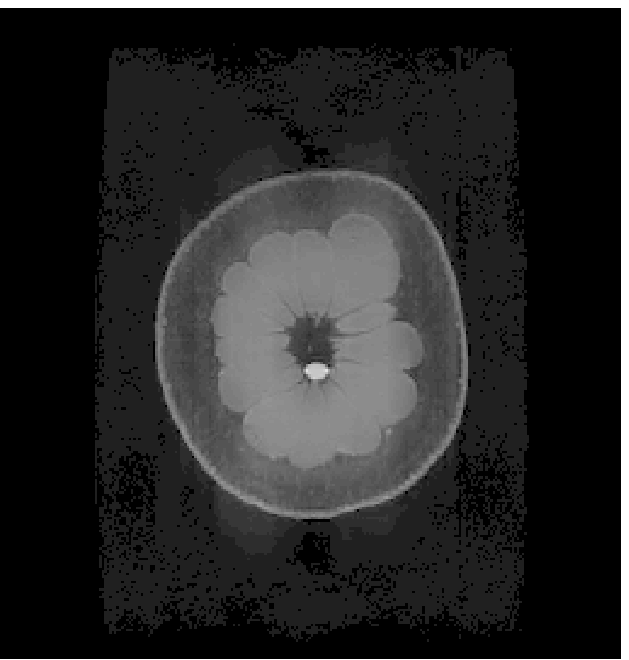}
\hspace{0.1mm}\includegraphics[width=3.3cm,angle=90,trim=0cm
5cm 0cm 5cm,
clip=true]{pomelo_non_adapative_k_space_4_fold.eps}
\end{minipage}\hspace{2mm}
\begin{minipage}{0.25\linewidth}
\includegraphics[width=82pt,trim=2.5cm
2.2cm 2.5cm 2.5cm,
clip=true]{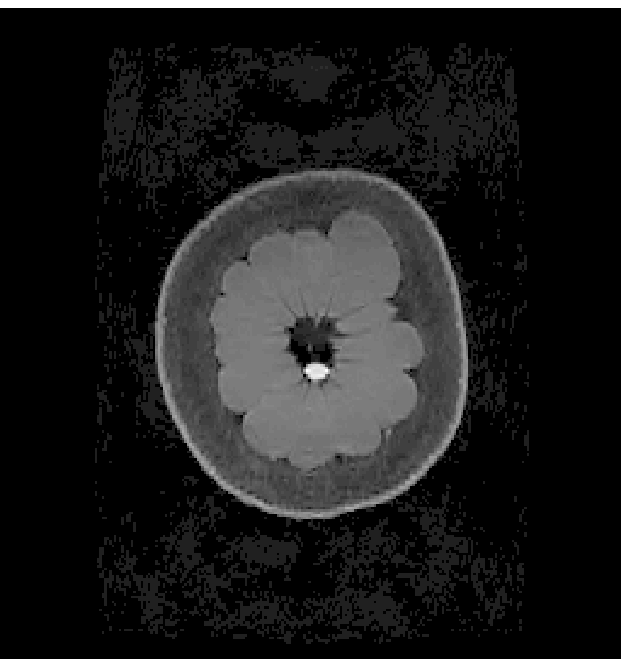}
\hspace{0.1mm}\includegraphics[width=3.3cm,angle=90,trim=0cm
5cm 0cm 5cm,
clip=true]{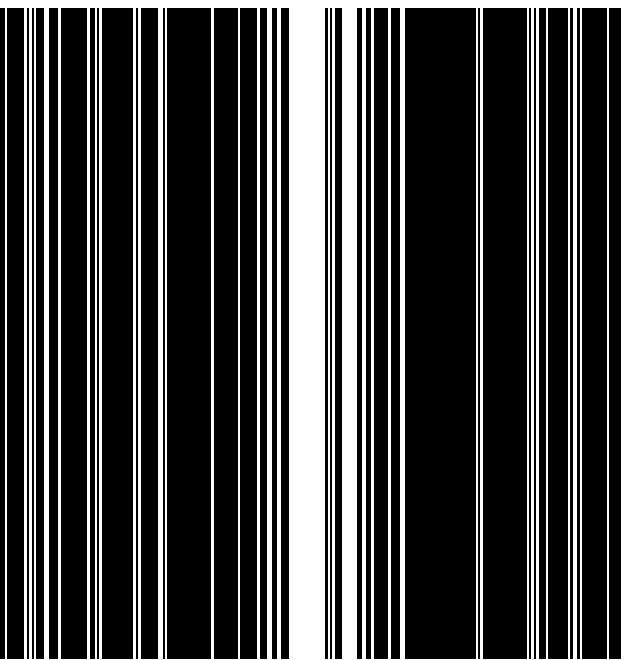}
\end{minipage}\\
\vspace{1mm}
\line(1,0){400}
\\\vspace{1mm}
\begin{minipage}{\linewidth}
\begin{minipage}{0.25\linewidth}
\vspace{-3cm}Nyquist sampled baseline scan:
 \end{minipage} \hspace{1mm}\includegraphics[width=82pt,trim=2.5cm
2.2cm 2.5cm 2.5cm,
clip=true]{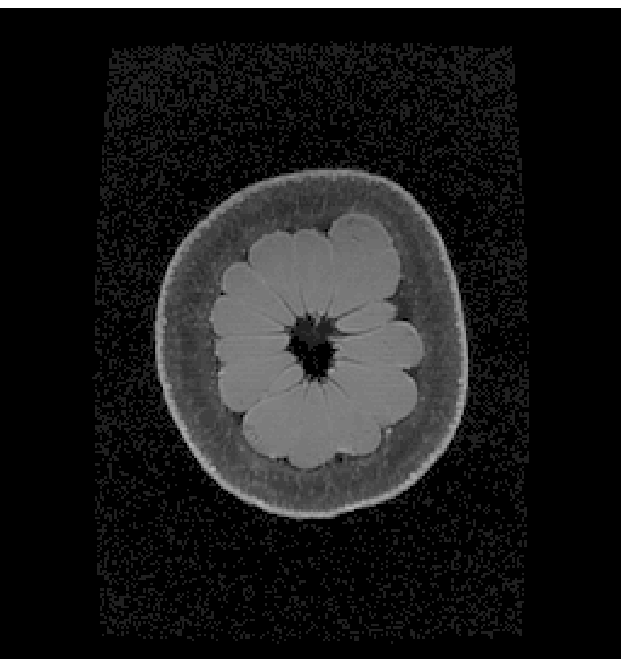}
 \hspace{2mm}
\begin{minipage}{0.25\linewidth}
\vspace{-3cm}Nyquist sampled follow-up scan:
 \end{minipage}
\includegraphics[width=82pt,trim=5.6cm
5.1cm 5.5cm 5.5cm,
clip=true]{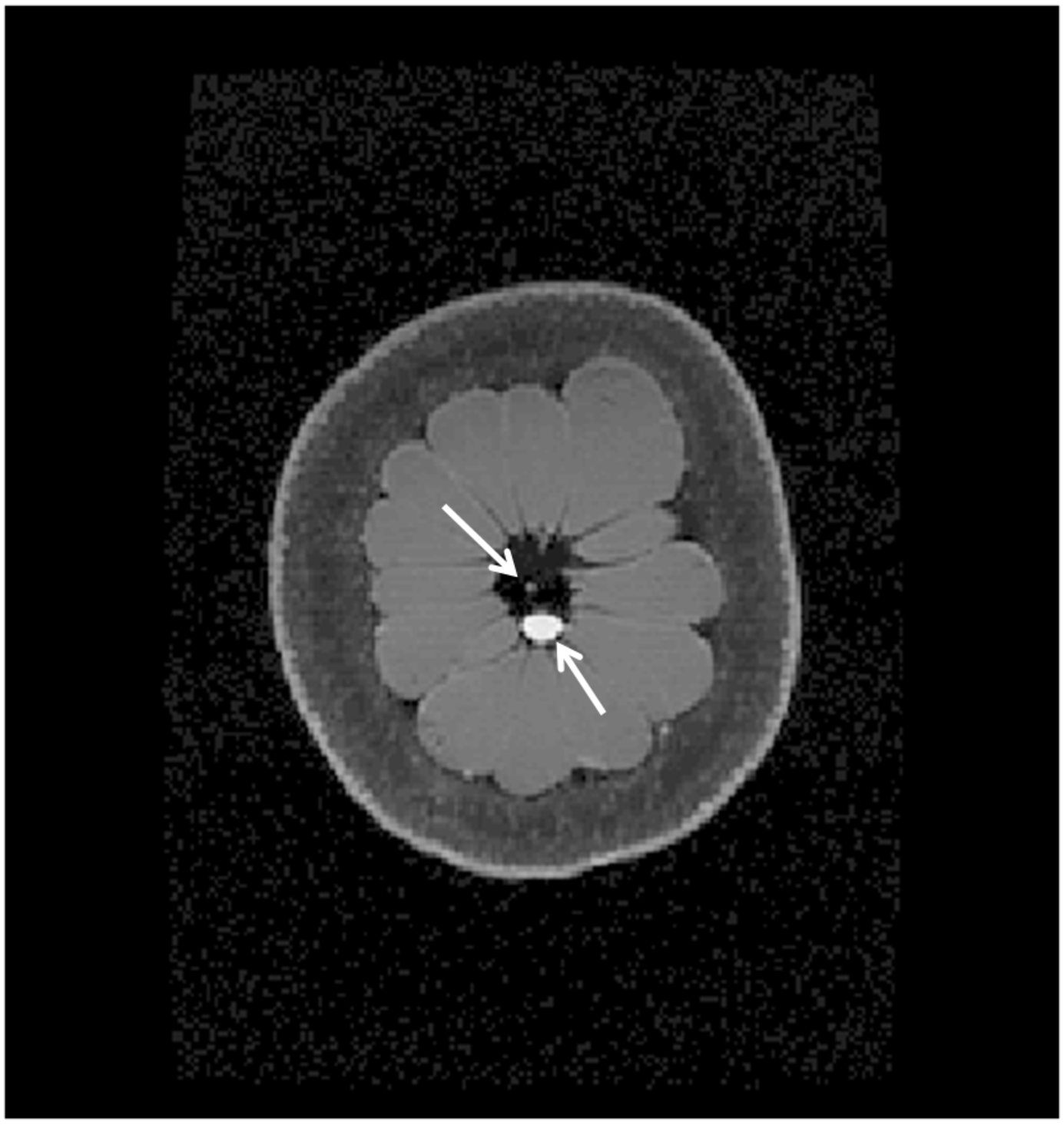}
\end{minipage}
\caption[Pomelo experiment results]{Pomelo experiment: The bottom row shows the scans of a pomelo before (left) and after (right) contrast agent injection, simulating a baseline and a follow-up scan. The major changes are seen in the follow-up scan as two enhancing regions, marked by arrows on the Nyquist sampled follow-up scan. Results of CS-MRI, TCS-MRI and LACS-MRI are presented at acceleration factors of 4, 6.4 and 10.6. The bar next to each image shows sampled phase encode lines under each scheme. It can be seen that LACS-MRI outperforms CS-MRI and TCS-MRI, and exhibits significantly improved resolution at acceleration factor of 4.}
\label{fig5}
\end{figure*}

The SNR values of the different reconstruction results are shown in Table \ref{tab1}. Figure \ref{fig5} shows the baseline and follow-up pomelo images and the reconstruction results. As expected, reconstruction results improve as the acceleration ratio decreases, for all methods. The major differences between the baseline and follow-up scans consist of two enhancing regions (marked in arrows), due to the injection of a contrast agent before the acquisition of the follow-up scan. 

In terms both image resolution and SNR, LACS-MRI outperforms TCS-MRI and CS-MRI, by exhibiting significantly improved recovery of the image at 4-fold acceleration, which also allows to identify changes versus the baseline scans, i.e. the two enhanced areas. Although both TCS-MRI and LACS-MRI utilize temporal similarity in the reconstruction process, this experiment emphasizes the advantage of embedding weighted-CS and adaptive sampling. Together with the weighting mechanism, the sampling locations (shown in the bar next to each image) which were chosen adaptively by LACS-MRI, lead to improved reconstruction results versus the pure random sampling used in CS-MRI and TCS-MRI.

\begin{center}
\begin{table}[!h]
\begin{center}
  \caption{Comparisons of the SNR (db), Pomelo experiment}
\begin{tabular}{|c|c|c|c| }

  \hline

      Acceleration factor  & CS-MRI & TCS-MRI & LACS-MRI   \\ \hline
      10.6 & $ 2.2468$ & $ 2.9052$ & $3.6107$ \\ \hline  
  6.4 & $2.9139$ & $ 7.0814$ & $ 13.8731$ \\ \hline
   4& $ 4.1401$&    $ 10.9464$&    $ 18.2111$ \\ \hline

\end{tabular}
 \label{tab1}
 \end{center}
\end{table}
\end{center}
\newpage
\begin{figure*}
\vspace{-3cm}
\begin{minipage}{\linewidth}\hspace{17mm}{ \parbox{3cm}{{ \singlespacing \footnotesize CS-MRI}}
\hspace{5mm}\parbox{3.5cm}{{\singlespacing \footnotesize TCS-MRI }}\hspace{0mm}\parbox{3.8cm}{{\singlespacing \footnotesize LACS-MRI }}}
\end{minipage}\\
\vspace{1mm}
\begin{minipage}{0.05\linewidth}
\vspace{-2cm}
\begin{turn}{90}{\hspace{1cm} $\times 16.6$  }\end{turn}
\end{minipage}
\includegraphics[width=100pt,trim=7cm
4.3cm 2.5cm 5cm,
clip=true]{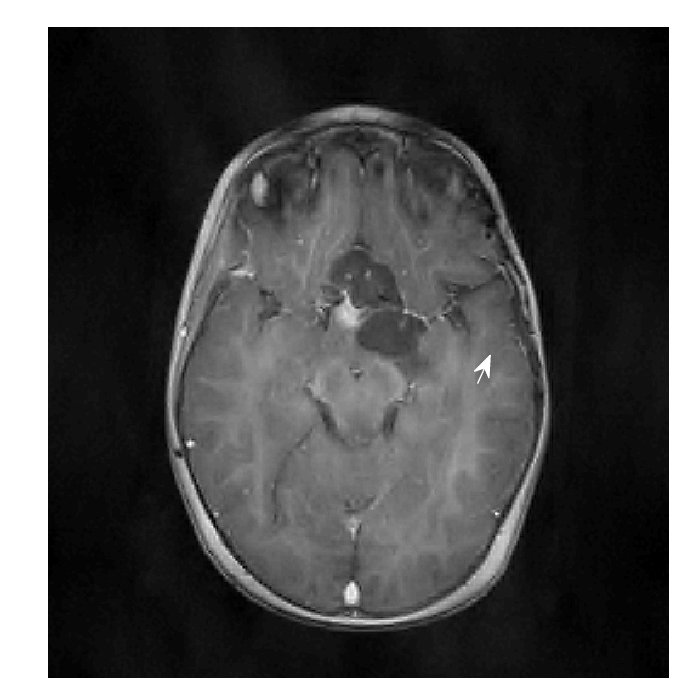}
\includegraphics[width=100pt,trim=7cm
4.3cm 2.5cm 5cm,
clip=true]{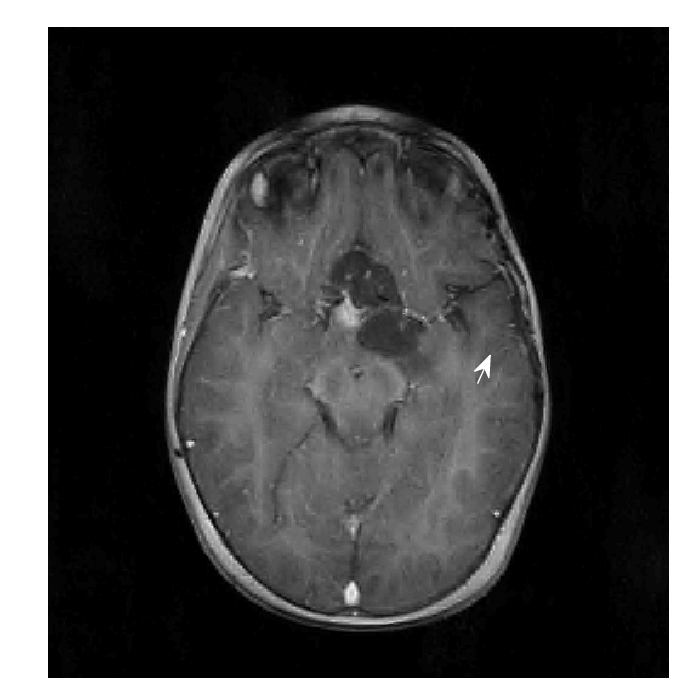}
\includegraphics[width=100pt,trim=7cm
4.3cm 2.5cm 5cm,
clip=true]{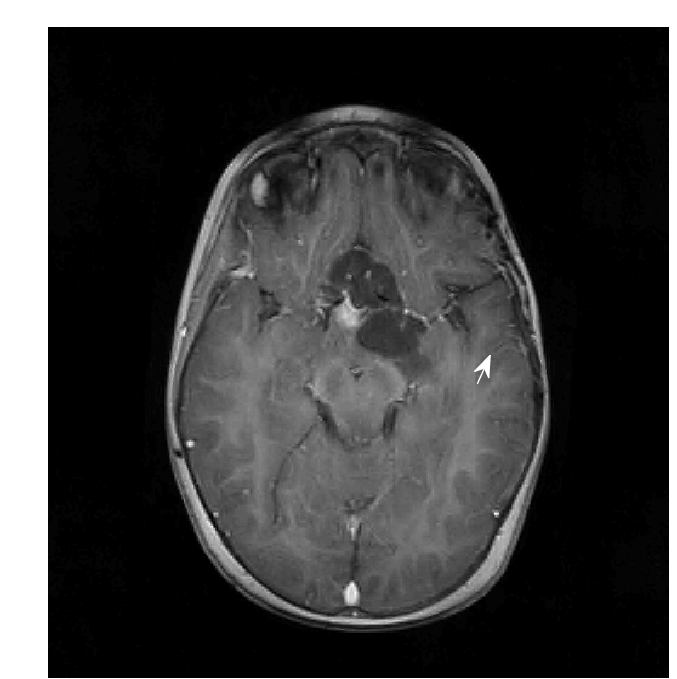}\\
\vspace{-1.7cm}
\begin{minipage}{0.05\linewidth}
\begin{turn}{90}{\hspace{3cm} $\times 10$  }\end{turn}
\end{minipage}
\includegraphics[width=100pt,trim=6.22cm
4.43cm 2.5cm 4.65cm,
clip=true]{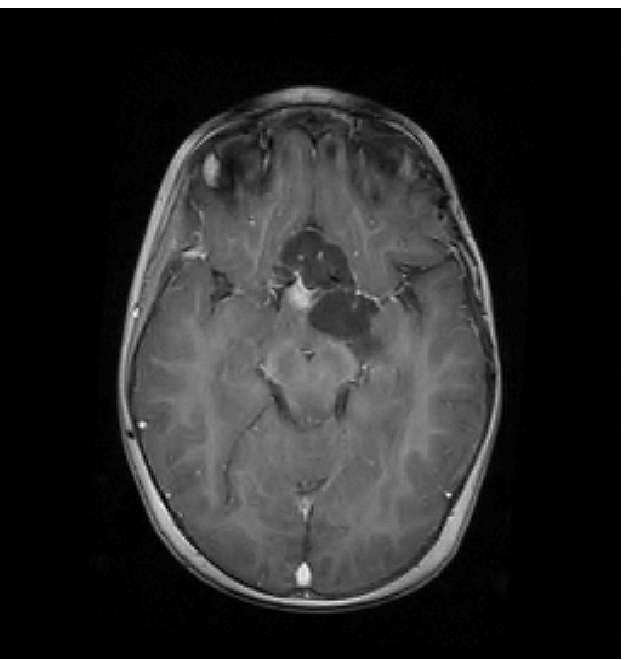}
\includegraphics[width=100pt,trim=6.22cm
4.43cm 2.5cm 4.65cm,
clip=true]{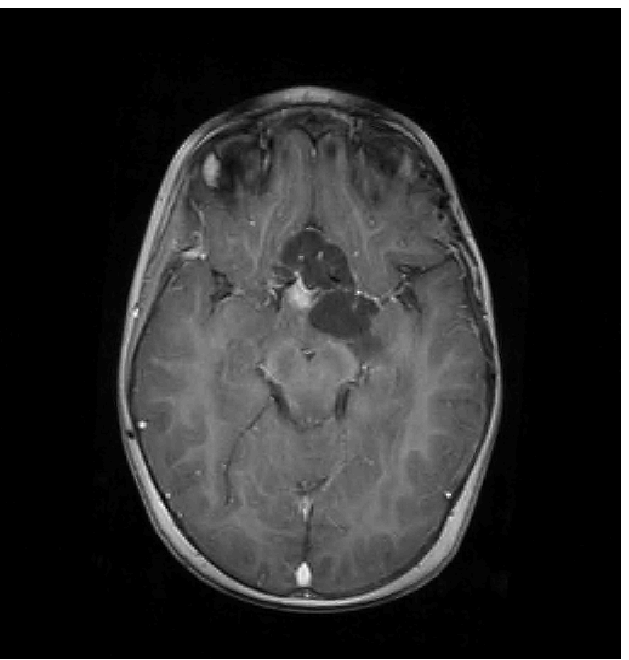}
\includegraphics[width=100pt,trim=6.22cm
4.43cm 2.5cm 4.65cm,
clip=true]{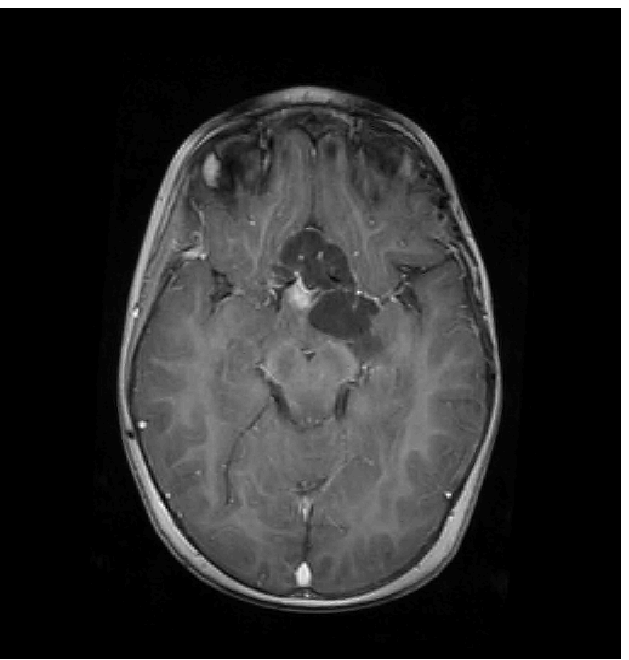}\\
\vspace{-1.2cm}
\begin{minipage}{0.05\linewidth}
\begin{turn}{90} {\hspace{3.5cm} $\times 5$  }\end{turn}
\end{minipage}
\includegraphics[width=100pt,trim=6.22cm
4.43cm 2.5cm 4.65cm,
clip=true]{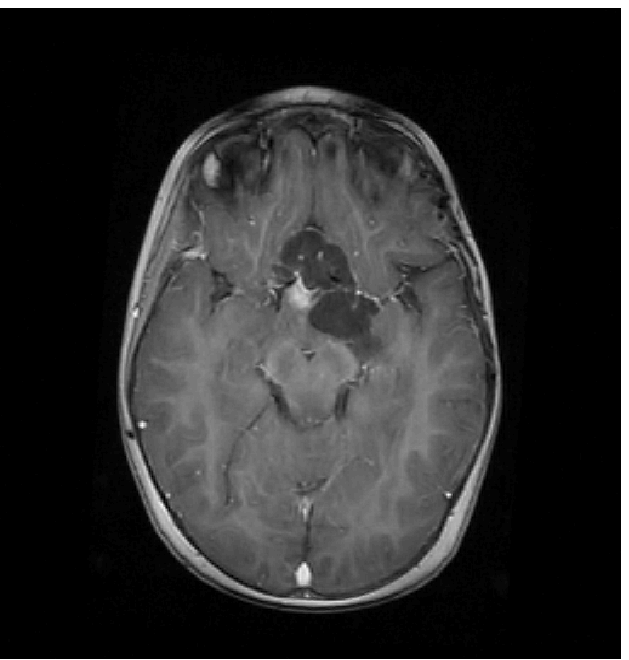}
\includegraphics[width=100pt,trim=6.22cm
4.43cm 2.5cm 4.65cm,
clip=true]{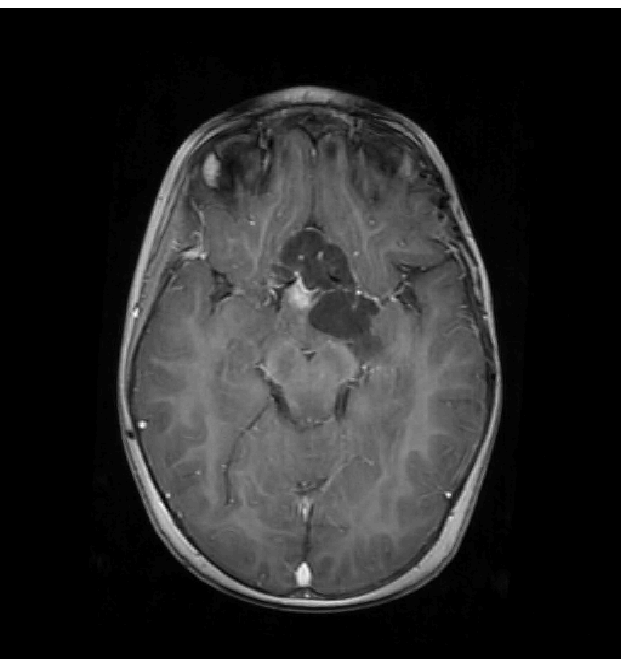}
\includegraphics[width=100pt,trim=6.22cm
4.43cm 2.5cm 4.65cm,
clip=true]{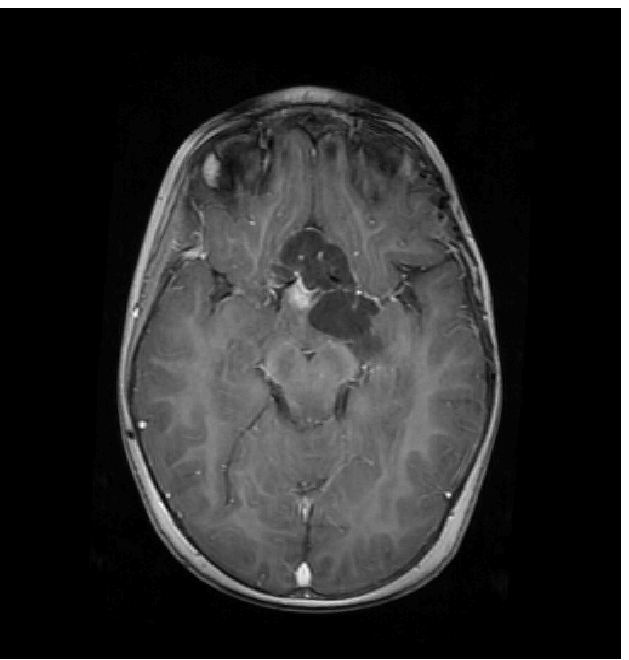}\\
\vspace{7mm}
\line(1,0){400}
\\\vspace{2mm}
\begin{minipage}{\linewidth}
\begin{minipage}{0.4\linewidth}
\hspace{00cm}
baseline scan ($\times 1$)\\
\hspace{1.6cm}
\includegraphics[width=100pt,trim=2cm
1cm 2cm 1.5cm,
clip=true]{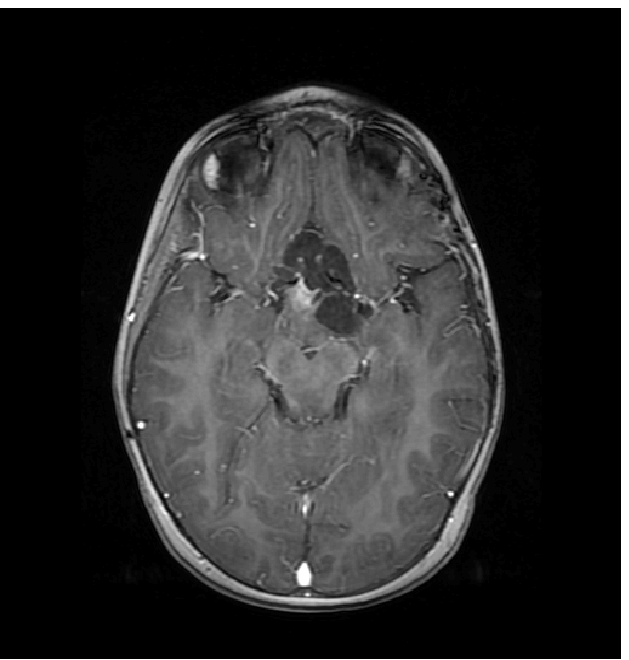}
 \end{minipage}
\begin{minipage}{0.59\linewidth}
\hspace{1.6cm}
follow-up scan ($\times 1$)\\
\begin{minipage}{3.5cm}
\includegraphics[width=100pt,trim=3.7cm
1cm 2cm 2.5cm,
clip=true]{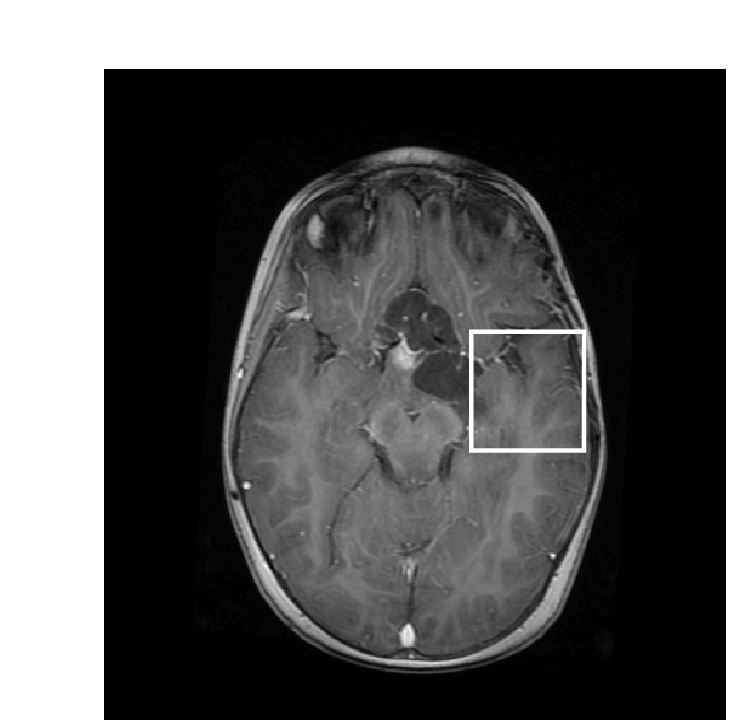}
 \end{minipage}
\begin{minipage}{3.4cm}
\includegraphics[width=100pt,trim=6.22cm
4.43cm 2.5cm 4.65cm,
clip=true]{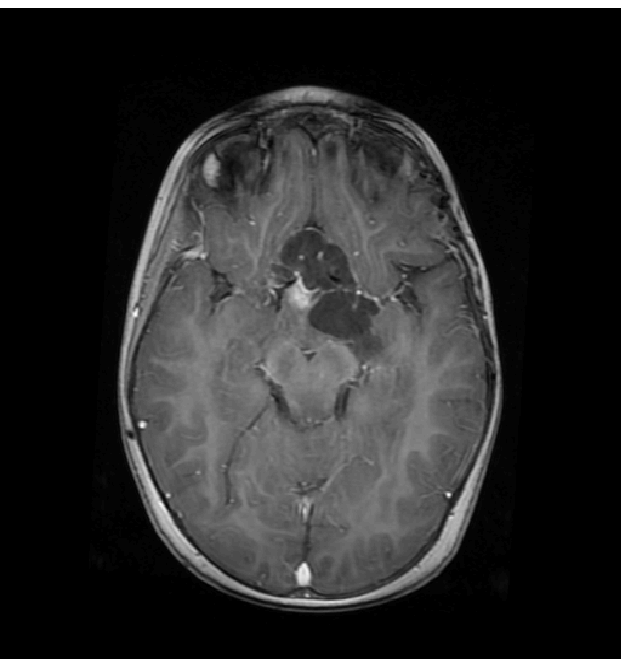}
 \end{minipage}
 \end{minipage}
\end{minipage}
\caption[3D FSPGR imaging results]{3D FSPGR results. While CS-MRI and TCS-MRI exhibit artifacts at 16.6-fold acceleration, LACS-MRI exhibits improved resolution (pointed by arrows). At 10-fold acceleration, TCS-MRI and LACS-MRI exhibit similar satisfactory performance, while CS-MRI exhibits some reconstruction artifacts. At 5-fold acceleration all method exhibit almost no loss of information with slightly better performance obtained by LACS-MRI. }
\label{fig8}
\end{figure*}

\subsection{3D Fast Spoiled Gradient Echo Brain Imaging }
Applying CS for 3D imaging allows undersampling in the 2D phase encode plane, thereby obtaining better performance than applying 2D CS slice by slice. We used two retrospectively acquired scans of a patient with OPG, where scans were acquired in an interval of six months. We used contrast enhanced 3D T1-weighted FSPGR sequence (matrix: $512\times 512 \times 46$, res = $0.47mm$, slice=$1.7mm$, TI/TR/TE=$450/14.5/6.3ms$, flip angle=$20^{\circ}$).

We rigidly registered the later scan to the former scan and then undersampled the  3DFT trajectory with corresponding acceleration factors of 5, 10 and 16.6 (20\%, 10\% and 6\% of the {\bf\emph k}-space). For LACS-MRI we begin with variable density random undersampling of 2\% of the k-space and acquire additional 2\% of the {\bf\emph k}-space at each iteration as described in Algorithm 1. 


Table \ref{tab2} shows the SNR values of the different reconstruction results. Figure \ref{fig8} shows the reconstruction results at 5,10 and 16.6-fold acceleration of a patient with OPG. Both TCS-MRI and LACS-MRI exhibit almost no loss of information at 10-fold acceleration. Similar results are obtained with CS-MRI only at 5-fold acceleration. 

This experiment shows that thanks to the ability to under-sample the 2D phase encode plane, the advantage of temporal similarity exploitation is emphasized. Therefore, LACS-MRI  allows shortening the scanning time by a factor of 10, with no significant loss of information in this case.

\begin{center}
\begin{table}[!h]
\begin{center}
  \caption{Comparisons of the SNR (db), 3D FSPGR experiment}
\begin{tabular}{|c|c|c|c| }

  \hline
        Acceleration factor  & CS-MRI & TCS-MRI & LACS-MRI   \\ \hline
  16.6 & $ 11.5093$ & $ 20.4286$ & $24.8350$ \\ \hline
  10 & $22.9917$ & $24.3437$ & $ 27.3621$ \\ \hline
       5& $ 25.4494$&    $28.0273$&    $ 29.9281$ \\ \hline

\end{tabular}
 \label{tab2}
 \end{center}
\end{table}
\end{center}

\subsection{2D Fast Spin-Echo Brain Imaging of Rapidly Changing Tumor}
To examine the performance of our approach with brain MRI data when the assumptions of similarity between consecutive scans is not valid, we used retrospectively acquired data of a patient with GBM. The patient was scanned twice within an interval of five months, and exhibited changes between scans that occupy more than 50\% of the brain region.  We used T2-weighted FSE sequence (matrix: $512\times 512$ res = $0.47mm$, 36 slices with $4mm$ thickness and no gap, TR/TE=$3500/113ms$, echo-train length=$24$, flip angle=$90^{\circ}$). We registered the follow-up scan to the baseline scan and examined the results of LACS-MRI ($N_k=16$), CS-MRI and TCS-MRI with acceleration factors of 4, 6.4, and 10.6.  

\begin{figure*}
\begin{minipage}{\linewidth}
\begin{minipage}{\linewidth}\hspace{2mm}\parbox{2.8cm}{ \singlespacing \footnotesize \centering Nyquist sampling (baseline)}\hspace{1mm}\parbox{2.7cm}{\centering \singlespacing \footnotesize Nyquist sampling (follow-up)}\hspace{1mm}{ \parbox{2cm}{\centering \singlespacing \footnotesize  CS-MRI (follow-up)}
\hspace{2mm}\parbox{2cm}{\centering \singlespacing \footnotesize TCS-MRI (follow-up) }\hspace{6mm}\parbox{2cm}{\centering \singlespacing \footnotesize LACS-MRI (follow-up)}\hspace{0mm}}
\end{minipage}
\begin{turn}{90}\parbox{4.0cm}{\hspace{-4mm} Reconstruction result  }\end{turn}\hspace{0.5mm}
\includegraphics[width=70pt,clip=true,trim=3.8cm
0.5cm 2cm 2.6cm]{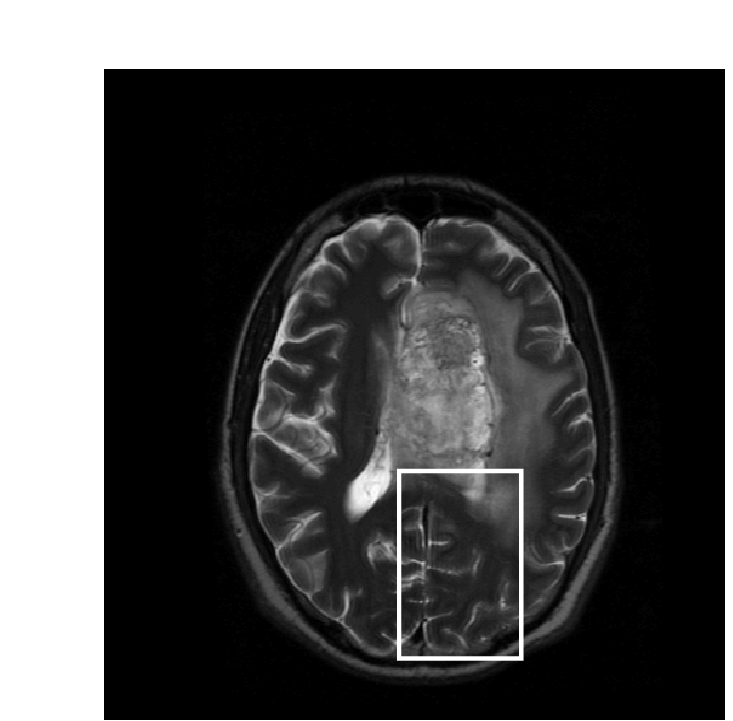}
\includegraphics[width=70pt,clip=true,trim=3.8cm
0.5cm 2cm 2.6cm]{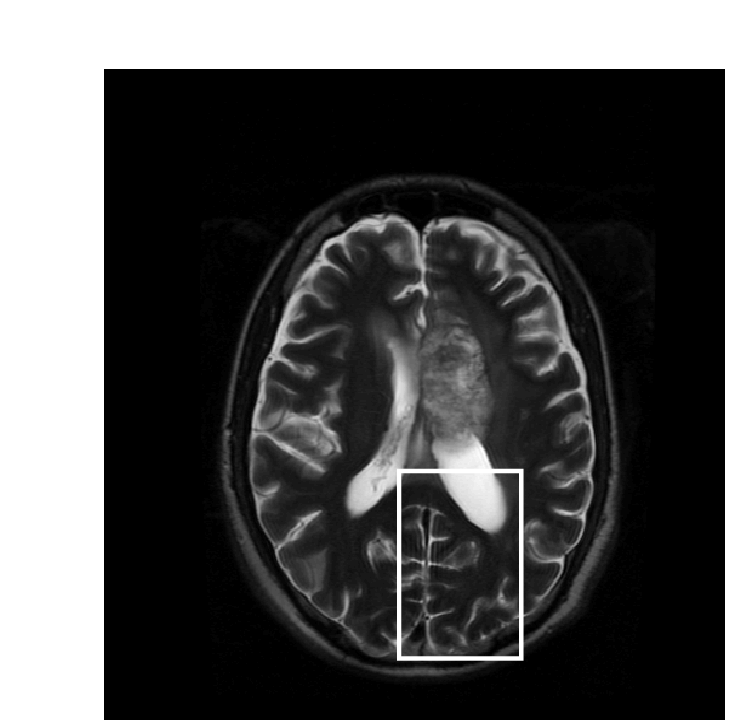}
\includegraphics[width=70pt,clip=true,trim=2cm
0.5cm 2cm 1.5cm]{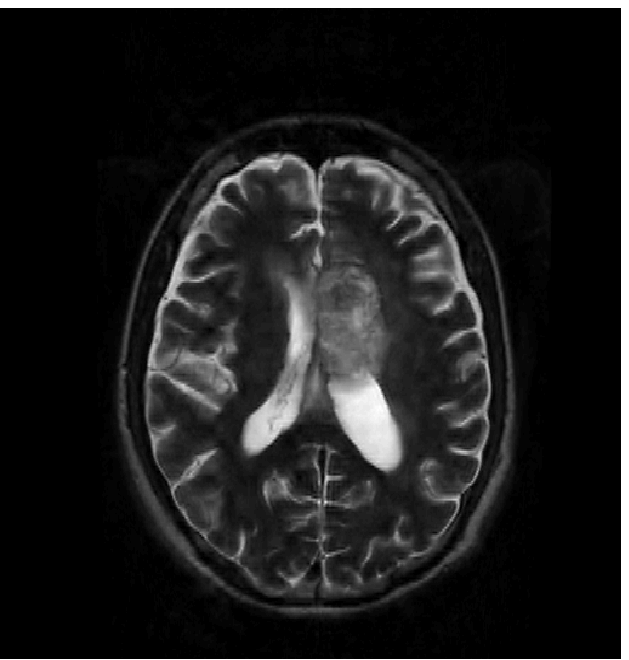}
\includegraphics[width=70pt,clip=true,trim=2cm
0.5cm 2cm 1.5cm]{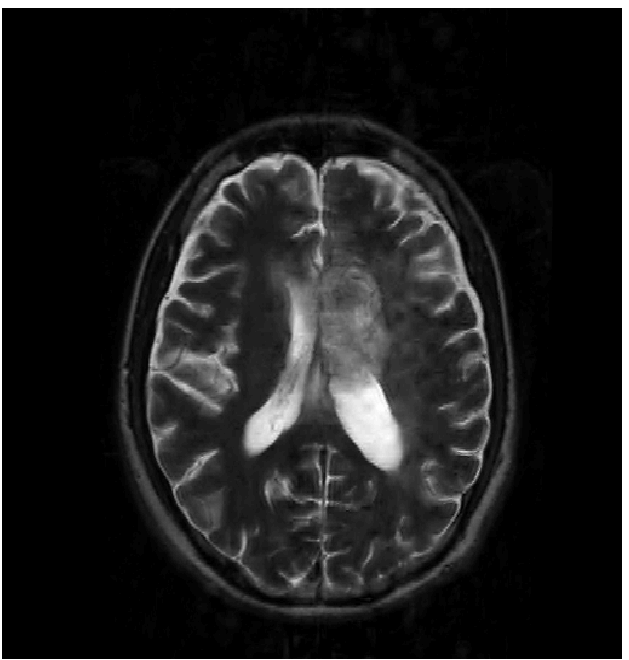}
\includegraphics[width=70pt,clip=true,trim=2cm
0.5cm 2cm 1.5cm]{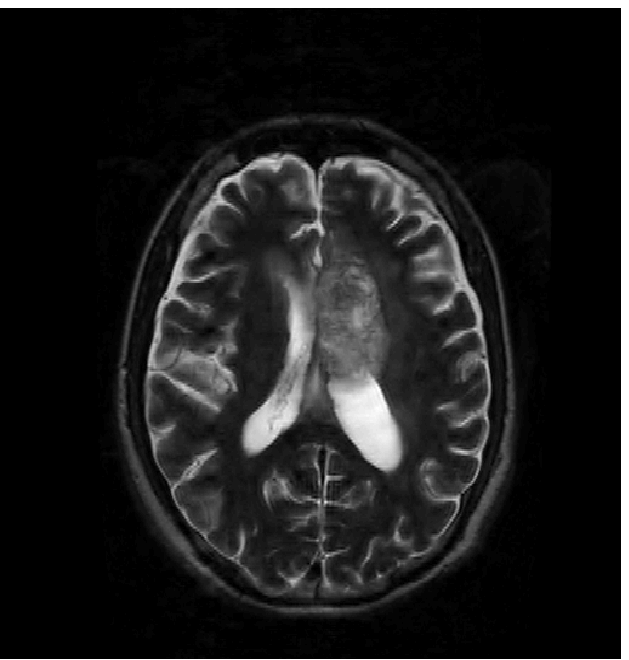}\\
\vspace{1mm}
\begin{turn}{90}\parbox{4cm}{\hspace{1mm}Difference image  }\end{turn}\hspace{0.5mm}
\includegraphics[width=70pt,clip=true,trim=2cm
0.5cm 2cm 1.5cm]{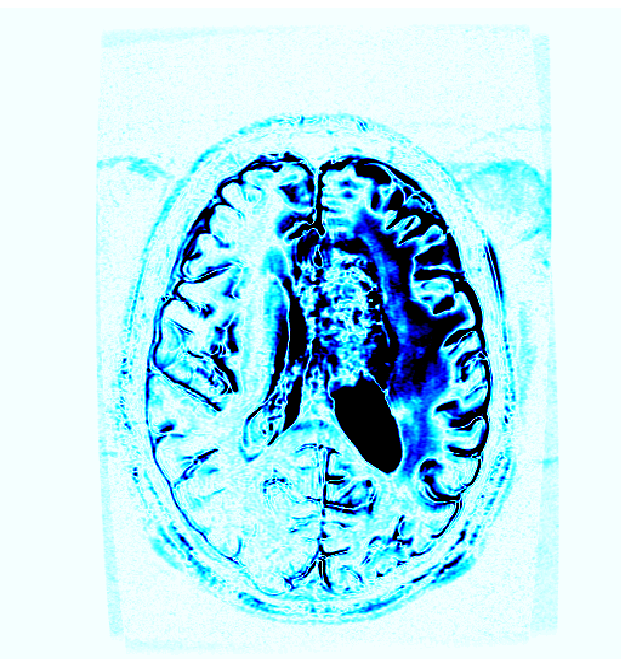}
\includegraphics[width=70pt,clip=true,trim=2cm
0.5cm 2cm 1.5cm]{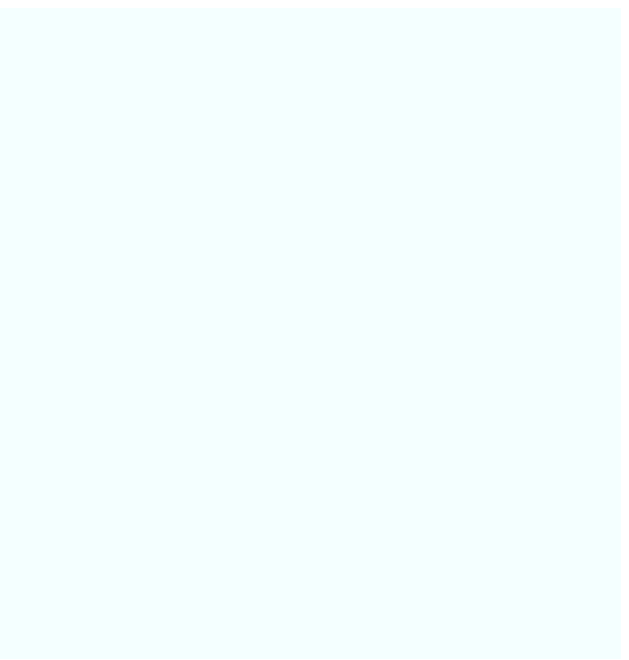}
\includegraphics[width=70pt,clip=true,trim=2cm
0.5cm 2cm 1.5cm]{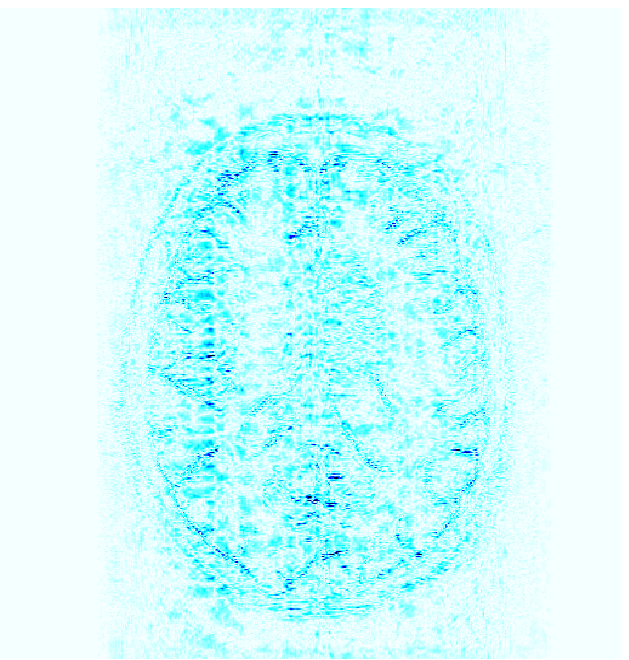}
\includegraphics[width=70pt,clip=true,trim=2cm
0.5cm 2cm 1.5cm]{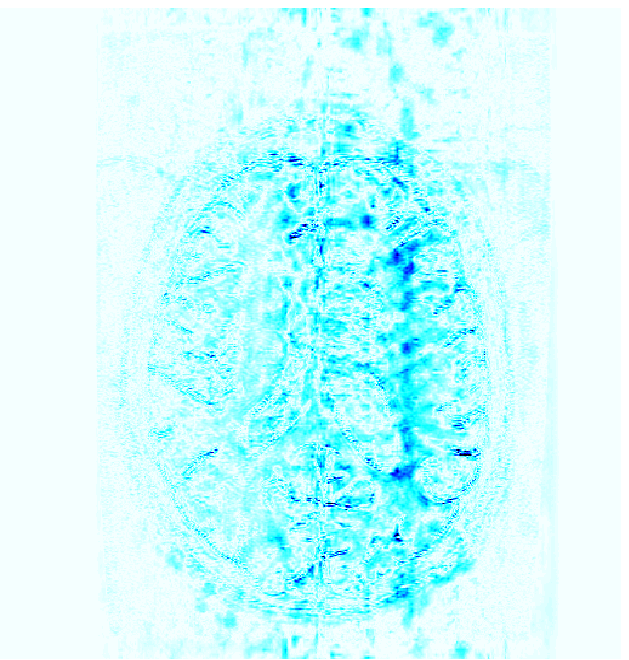}
\includegraphics[width=70pt,clip=true,trim=2cm
0.5cm 2cm 1.5cm]{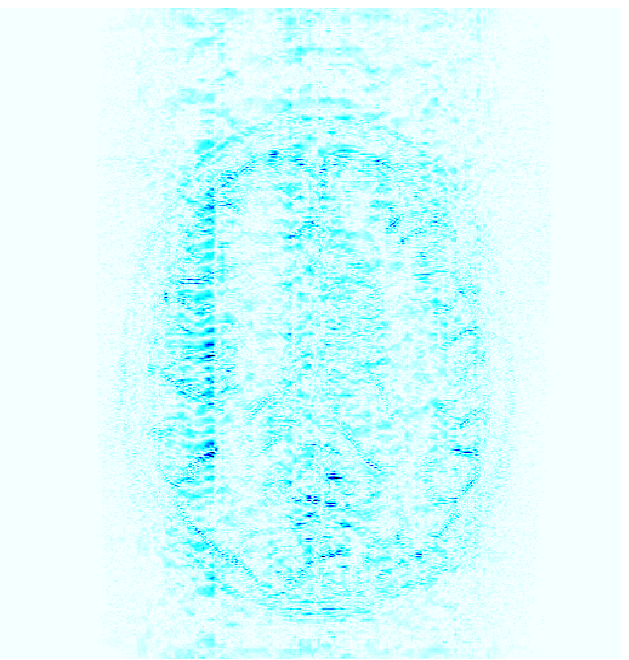}
\\
\begin{turn}{90}\parbox{4.2cm}{\hspace{0mm}Reconstruction (zoom)  }\end{turn}\hspace{0.5mm}
\includegraphics[width=70pt,clip=true,trim=5.02cm
1cm 3.5cm 6.95cm]{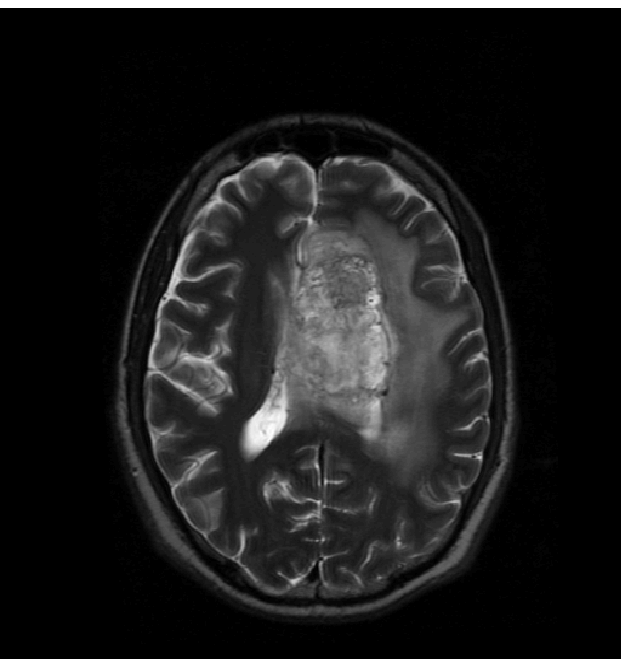}
\includegraphics[width=70pt,clip=true,trim=6cm
1cm 3.5cm 7cm]{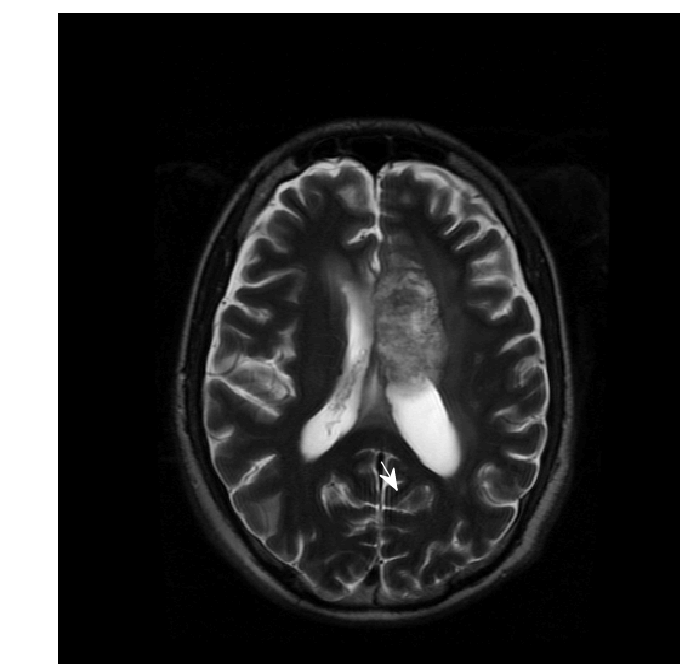}
\includegraphics[width=70pt,clip=true,trim=6cm
1cm 3.5cm 7cm]{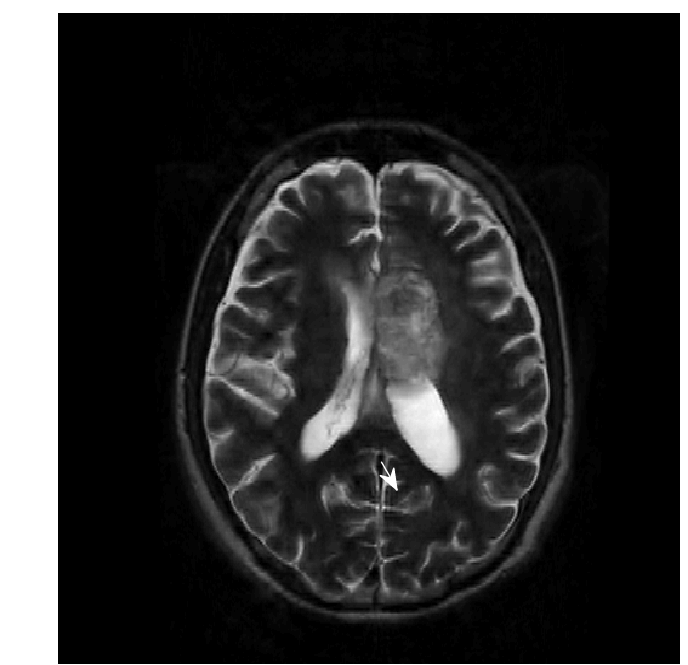}
\includegraphics[width=70pt,clip=true,trim=6cm
1cm 3.5cm 7.05cm]{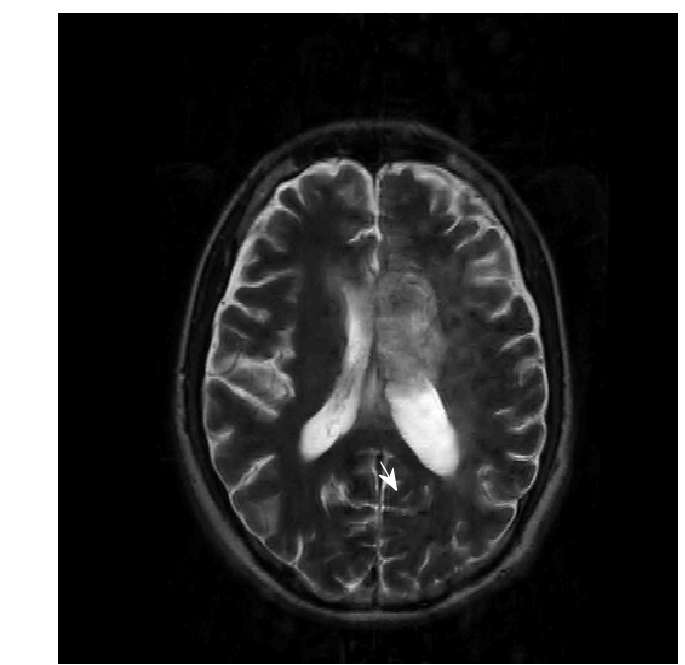}
\includegraphics[width=70pt,clip=true,trim=6cm
1cm 3.5cm 7.05cm]{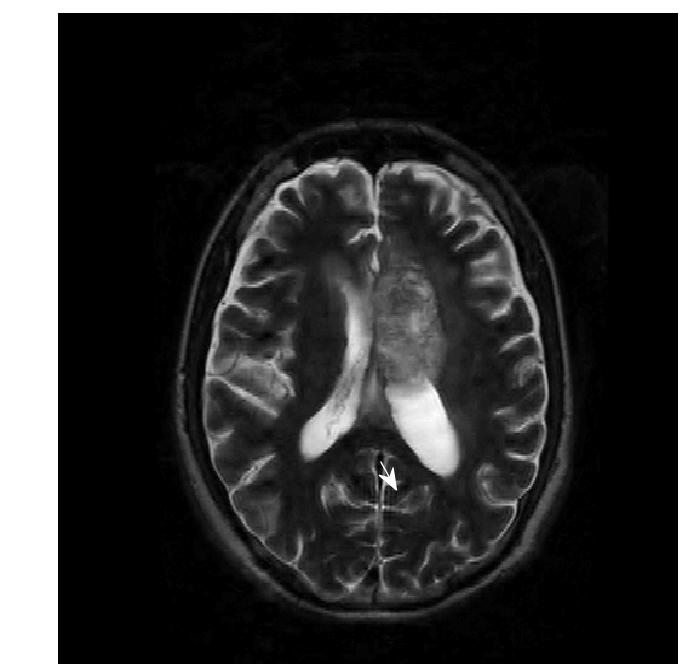}
\\
\end{minipage}
\caption[2D FSE brain imaging results when similarity between baseline and follow-up is minimal]{2D FSE brain imaging results when similarity between baseline and follow-up is minimal. The leftmost column shows the baseline scan used as $\vec{x}_0$ for TCS-MRI and LACS-MRI. The difference image between the baseline and the follow-up scans shows substantial changes due to rapid tumor changes. The last row shows an enlargement of the white rectangle. LACS-MRI and CS-MRI exhibit similar results, where TCS-MRI exhibits sub-optimal results in the changes regions, due to its wrong assumption of substantial between the baseline and the follow-up scans. }
\label{fig7}
\end{figure*}

Table \ref{tab3} shows the SNR values for different reconstructions and Figure \ref{fig7} shows reconstruction results visually,  at acceleration factor of 4 (25\% of the {\bf\emph k}-space). In this case, there are major changes between the baseline and the follow-up scans due to therapy response. As a result, TCS-MRI exhibits poor performance in the vicinity of the changing tumor, since it is partially based on similarity between the consecutive scans, an assumption which is not valid in this case. 

LACS-MRI, however, convergences to a result which is similar to CS-MRI. This is obtained thanks to the adaptive sampling and the weighting mechanism embedded in  LACS-MRI, which reduces the weight given to the similarity to prior scan in the reconstruction process, if such a similarity does not exist.

\begin{center}
\begin{table}[!h]
\begin{center}
  \caption{Comparisons of the SNR (db), 2D FSE experiment, major temporal changes changes.}
\begin{tabular}{|c|c|c|c| }

  \hline
        Acceleration factor  & CS-MRI & TCS-MRI & LACS-MRI   \\ \hline
        10.6 & $ 8.2324$ & $ 7.4092$ & $8.9705$ \\ \hline
  6.4 & $9.1760$ & $ 8.2584$ & $ 9.7180$ \\ \hline
 4& $ 20.1972$&    $18.1775$&    $ 20.6306$ \\ \hline
\end{tabular}
 \label{tab3}
 \end{center}
\end{table}
\end{center}

\section{Discussion}
\label{discussion}

\subsection{Reproducing image orientation in longitudinal studies}
\label{applicn ability_section}
The exploitation of temporal similarity in longitudinal studies assumes that the past scan and the follow-up scan being acquired are spatially matched. In our experiments, we worked with retrospectively acquired data and rigidly registered the later scan to the former scan. Note that our method exhibits reliable performance even when minor registration errors exist (as can be seen in the difference images in Fig. \ref{fig7}), thanks to the similarity of adjacent pixels in MRI.

The practical implementation of our method for prospective acquisition of follow-up scans requires reproducing the past scan's slice positions for the scan being acquired. This spatial matching is currently offered as a feature by some MRI vendors \citep{signapulse1}, and is based on several anatomical landmark. As previously noted, our method is expected to overcome minor matching errors which might be produced by these tools thanks to fact that adjacent pixels have similar values in many MRI applications.

An additional practical solution could utilize the adaptive sampling mechanism. Partially sampled data can be used to determine the spatial orientation of the patient vs. the reference image, during acquisition. This approach have been tested to compensate motions during  imaging \citep{lingala2011unified,feng2013golden} and can be utilized for the longitudinal scanning case.

The applicability of the proposed method to other regions outside of the brain, such as abdominal MRI is more complicated as it may require non-rigid registration. Therefore, this paper focuses on the application of CS for longitudinal brain studies, while the implementation for regions outside of the brain is left for future research.

\subsection{Grey Level Intensity Differences}
The exploitation of temporal similarity is performed by utilizing the fact that the difference between the baseline and the follow-up scans is sparse, and consists changes in anatomy or pathology. However, changes between baseline and follow-up scans may be the result of other factors, such as acquisition parameters and field inhomogeneity. In this work, we normalized the grey level intensity values of the scans to match the same scale, in order to minimize the effect of external factors on changes between the scans. This normalization was sufficient for producing the results presenting in this paper. In prospective scanning, the normalization coefficients can be determined iteratively, during the sampling and reconstruction process.

In the special case of longitudinal studies, scans are in many cases acquired in the same scanning site with the same scanning protocol, to minimize the effect of external parameters on the resulted clinical follow-up. When the baseline and follow-up scans are acquired with different acquisition parameters on different systems, we can still exploit the structural similarity between the baseline and the follow-up scans for rapid acquisition. While this extension is not in the scope of this paper, the reader is referred to the works of Bilgic et al. \citep{bilgic2011multi} and Huang et al. \citep{huang2012fast} who utilize the structural similarity between various MRI sequences for rapid scanning.

\subsection{Computational Complexity}
Fast algorithms for solving  $\ell_1$ minimization problems have gained much attention recently. Besides FISTA \citep{beck2009fast}, used in our experiments, we find NESTA \citep{becker2011nesta}, SALSA \citep{afonso2010fast}, C-SALSA \citep{afonso2011augmented}, SPGL1 \citep{van2008probing} among other approaches that report improved convergence time and have not been explored in this work.

Fast convergence time is important in our adaptive sampling approach, that requires solving $\ell_1$ minimization problem to decide on the sampling locations for the next iteration. In a Matlab (The MathWorks, Natick, MA) implementation, each $\ell_1$ minimization for $256\times 256$ brain image requires approximately 35 seconds. 

While our implementation of FISTA at present does not attain the high rates required for solving $\ell_1$ minimization for real-time applications, we expect a significant reduction in the reconstruction time by code optimization. Examining new approaches for $\ell_1$ minimization, algorithmic simplifications, combined with massively parallel digital computation could allow our framework to be used in the future in order to allow the adaptive sampling mechanism operate during MRI scanning.  

\section{Conclusions}
\label{conclusions}
Repeated scans constitute a substantial portion of MRI scanning today, mainly to track changes in pathologies and to monitor treatment efficacy. We presented LACS MRI based on adaptive sampling and weighted-CS for rapid MR imaging of longitudinal studies. We demonstrated
experimental verification of several implementations for
2D and 3D Cartesian imaging. We showed that the temporal sparsity
of longitudinal MR images can be exploited to significantly reduce
scan time of follow-up scans, or alternatively, improve their resolution. 

We demonstrated that unlike other CS-MRI applications, sparsity in the temporal domain is not guaranteed in longitudinal studies. We showed that our method provides almost no loss of information at 10-fold acceleration of 3D brain scans, when there is substantial similarity between baseline and follow-up scan. Moreover, our method adapts to a scenario in which there is substantial difference between the scans and results converge to state-of-the-art CS-MRI in this case. 

LACS-MRI can play a major part in applications that consist of patient's disease follow-up and changes monitoring. This could be a first step towards utilizing the huge amount of data in picture archiving and communication systems (PACS) to speed-up MRI. 
\label{conclusions}

\appendix
\section{Fast Iterative Shrinkage-Thresholding Algorithm for LACS-MRI}
\label{app_a}
LACS-MRI poses an unconstrained problem in so-called Lagrangian form:
\begin{equation*}
\begin{aligned}
& \underset{\vec{x}}{\text{min}}
&  \|\mat{F}_u\vec{x}-\vec{y}\|_2^2+\lambda_1\|\mat{W}_1\mat{\Psi}\vec{x}\|_1+\lambda_2\|\mat{W}_2(\vec{x}-\vec{x}_0)\|_1.
\end{aligned}\tag*{(A1)}
\label{eqA1}
\end{equation*}
\noindent We solve \ref{eqA1} with Fast Iterative Shrinkage-Thresholding (FISTA) \citep{beck2009fast}. Originally, FISTA was introduced to solve the following problem:
\begin{equation*}
\begin{aligned}
& \underset{\vec{x}}{\text{min}}
&  \|\mat{A}\vec{x}-\vec{y}\|_2^2+\lambda\|\vec{x}\|_1.
\end{aligned}\tag*{(A2)}
\label{eqA2}
\end{equation*}

Recently, Tan et al. \citep{tan2014smoothing} proposed a FISTA-based that allows for weighted-$\ell_1$ of the form:  
\begin{equation*}
\begin{aligned}
& \underset{\vec{x}}{\text{min}}
&  \|\mat{A}\vec{x}-\vec{y}\|_2^2+\lambda\|\mat{D}\text{*}\vec{x}\|_1
\end{aligned}\tag*{(A3)}
\label{eqA3}
\end{equation*}
\noindent Their algorithm, coined SFISTA, is described as Algorithm 1 in \citep{tan2014smoothing}. Based on \citep{tan2014smoothing} we define the following:
\begin{equation*}
\begin{aligned}
f(\vec{x})=\mat{F}_u\vec{x}-\vec{y}
\end{aligned}\tag*{(A4)}
\label{eqA4}
\end{equation*}
\begin{equation*}
\begin{aligned}
&g_{1\mu}(\mat{W}_1\mat{\Psi}\vec{x})=
& \underset{\vec{u}}{\text{min}}\{\lambda_1\|\vec{u}\|_1+\frac{1}{2\mu}\|\vec{u}-\mat{W}_1\mat{\Psi}\vec{x}\|_2^2\}
\end{aligned}\tag*{(A5)}
\label{eqA5}
\end{equation*}
\begin{equation*}
\begin{aligned}
&g_{2\mu}(\mat{W}_2(\vec{x}-\vec{x}_0))=
& \underset{\vec{u}}{\text{min}}\{\lambda_2\|\vec{u}\|_1+\frac{1}{2\mu}\|\vec{u}-\mat{W}_2(\vec{x}-\vec{x}_0)\|_2^2\}
\end{aligned}\tag*{(A6)}
\label{eqA6}
\end{equation*}

Extending SFISTA to solve \ref{eqA1} results in the following algorithm:
\begin{algorithm}[H]{\bf{SFISTA algorithm for LACS-MRI}}
\begin{algorithmic}
\REQUIRE \hspace{3mm} \\
{\bf\emph k}-space measurements: $\vec{y}$; \\
Sparsifying transform operator: $\mat{\Psi}$\\
An $N\times N$ undersampling operator in the k-space domain: $\mat{F}_u$\\
Image from earlier time-point: $\vec{x}_0$;\\
Tuning constants: $\lambda_1,\lambda_2,\mu$\\
An upper bound: $L\geq \|\mat{F}_u\|_2^2+\frac{\|\mat{W_1\Psi}\|_2^2+\|\mat{W_2}\|_2^2}{\mu}$
\ENSURE Estimated image: $\mat{\hat{x}}$
\renewcommand{\algorithmicrequire}{\textbf{Initialize:}}
 \REQUIRE \hspace{3mm} \\
\STATE $\vec{x}_0=\vec{z}_1=\mat{F}_u^*\vec{y}$, $t_1=1$
\renewcommand{\algorithmicrequire}{\textbf{Iterations:}}
 \REQUIRE \hspace{3mm} \\
\STATE {\bf Step k:} $(k\geq 1)$ Compute
\STATE $\nabla f(\vec{z}_k)= \mat{F}_u^*(\mat{F_u}\vec{z}_k-\vec{y})$
\STATE $\nabla g_{1\mu}(\mat{W}_1\mat{\Psi}\vec{x}_{k-1})=\frac{1}{\mu}\mat{W}_1\mat{\Psi^*}(\mat{W}_1\mat{\Psi}\vec{x}_{k-1}-\Gamma_{\lambda_1\mu}\left(\mat{W}_1\mat{\Psi}\vec{x}_{k-1})\right)$
\STATE $\nabla g_{2\mu}(\mat{W}_2(\vec{x}_{k-1}-\vec{x}_0))=\frac{1}{\mu}\mat{W}_2(\mat{W}_2(\vec{x}_{k-1}-\vec{x}_0)-\Gamma_{\lambda_2\mu}\left(\mat{W}_2(\vec{x}_{k-1}-\vec{x}_0)\right)$
\STATE $\vec{u}_k=\vec{z}_k-\frac{1}{L}\left(\nabla f(\vec{z}_k)+\nabla g_{1\mu}(\mat{W}_1\mat{\Psi}\vec{x}_{k-1})+\nabla g_{2\mu}(\mat{W}_2(\vec{x}_{k-1}-\vec{x}_0))\right)$
\STATE $t_{k+1}=\frac{1+\sqrt{1+4t_k^2}}{2}$
\STATE $\vec{x}_k=argmin\{H_{\mu}(\vec{x}):\vec{x}=\mat{u}_k,\mat{x}_{k-1}\}$
\STATE $\vec{z}_{k+1}=\vec{x}_k+\frac{t_k}{t_{k+1}}\left(\vec{u}_k-\vec{x}_k\right)+\frac{t_k-1}{t_{k+1}}\left(\vec{x}_k-\vec{x}_{k-1}\right)      $
\end{algorithmic}
\end{algorithm}
\noindent where the notation $\|\cdot\|_2$ for matrices denotes the largest singular value, and   $\Gamma_{\lambda \mu }(\vec{z})$ is the soft shrinkage operator, which operates element-wise on $\vec{z}$ and defined as (for complex valued $z_i$):
\begin{equation*}
 \Gamma_{\lambda \mu }(z_i)=\begin{cases}
    \frac{|z_i|-\lambda\mu}{|z_i|}z_i, & |z_i|>\lambda\mu  \\
   0, & otherwise
  \end{cases}
\tag*{(A7)}
\label{eqA7}
\end{equation*}
and:
\begin{equation*}
\begin{aligned}
H_{\mu}(\vec{x})=\|\mat{F}_u\vec{x}-\vec{y}\|_2^2+g_{1\mu}(\mat{W}_1\mat{\Psi}\vec{x})+g_{2\mu}(\mat{W}_2(\vec{x}-\vec{x}_0)) 
\end{aligned}\tag*{(A8)}
\label{eqA8}
\end{equation*}

The algorithm above minimizes  \ref{eqA1}, where the trade-off between the two sparsity assumptions is controlled by the ratio between $\lambda_1$ and $\lambda_2$, via $\Gamma(\cdot)$, and the overall convergence is controlled by $\mu$. Typical values for $\lambda_1$ and $\lambda_2$ are in the range of $[0.001,0.9]$. The value of $\mu$ usually depends with the values of $\lambda_1$ and $\lambda_2$ and is in the range of $[10^{-9},100]$. The value of $\mu=10^{-3}(\frac{\lambda_1+\lambda_2}{2})^{-1}$ was used in our experiments. The number of iterations varies with different objects, problem size, accuracy and undersampling. Examples in this paper required between 30 and 50 iterations.

\section*{{A}cknowledgements}
The authors wish to thank the Gilbert Israeli Neurofibromatosis Center (GINFC) for
providing the real data and supporting the medical part of the paper. This research was
supported by the Ministry of Science and Technology, Israel.
\section*{{R}eferences}
\bibliographystyle{model2-names}
\bibliography{paper_for_media}







\end{document}